\documentclass[reprint,superscriptaddress]{revtex4}
\usepackage{url}
\usepackage{amsmath}
\usepackage{mathtools}
\usepackage{bbold}
\usepackage{mathrsfs}

\usepackage{graphicx}
\usepackage[caption=false]{subfig}
\usepackage{epstopdf} 
\usepackage{wrapfig}
\usepackage{braket}
\usepackage{xcolor}
\usepackage{comment}
\usepackage{lipsum}
\usepackage{cancel}
\usepackage{hyperref}

\newcommand{\RNum}[1]{\uppercase\expandafter{\romannumeral #1\relax}}

\newcommand{\btheta}{\boldsymbol{\theta}}

\newcommand{\step}{{\mathrm{step}}}
\newcommand{\diag}{{\mathrm{diag}}}
\newcommand{\onehalf}{\frac{1}{2}}

\def\ket#1{| #1\rangle}

\def\ud{\mathrm{d}}

\newcommand{\nep}{\mathrm{e}}

\newcommand{\dQA}{\mathrm{\scriptscriptstyle dQA}}
\newcommand{\QAOA}{\mathrm{\scriptscriptstyle QAOA}}

\newcommand{\A}{{\mathbf A}}
\newcommand{\B}{{\mathbf B}}
\newcommand{\C}{{\mathbf C}}
\newcommand{\rmC}{{\mathrm C}}
\newcommand{\G}{{\mathbf G}}

\newcommand{\Heis}{\mathrm{\scriptscriptstyle H}}

\newcommand{\transpose}{\mathsf{T}}

\newcommand{\Nbasis}{\mathrm{N_c}}
\newcommand{\F}{{\mathbf F}}
\newcommand{\bfD}{{\mathbf D}}
\newcommand{\U}{{\mathbf U}}
\newcommand{\V}{{\mathbf V}}

\newcommand{\prm}{\mathrm{p}}

\newcommand{\opc}[1]{{\hat{c}^{\phantom \dagger}}_{#1}}
\newcommand{\opcdag}[1]{{\hat{c}^{\dagger}}_{#1}}

\newcommand{\PauliSigma}{\hat{\sigma}}

\newcommand{\identity}{\mathbf{1}}

\newcommand{\Ho}{\hat{H}}
\newcommand{\Uo}{\hat{U}}

\newcommand{\Nsites}{\mathrm{N}}
\newcommand{\Nfermions}{\mathrm{\hat{N}_F}}
\newcommand{\res}{\mathrm{res}}
\newcommand{\fin}{\mathrm{fin}}
\newcommand{\eff}{\mathrm{eff}}

\newcommand{\Ptrot}{\mathrm{P}}

\newcommand{\mathH}{\mathbb{H}}
\newcommand{\opgamma}[1]{{\hat{\gamma}^{\phantom \dagger}}_{#1}}
\newcommand{\opgammadag}[1]{{\hat{\gamma}^{\dagger}}_{#1}}
\newcommand{\opgammatilde}[1]{{\tilde{\gamma}^{\phantom \dagger}}_{#1}}
\newcommand{\opgammatildedag}[1]{{\tilde{\gamma}^{\dagger}}_{#1}}

\newcommand{\opbfPsi}[1]{{\widehat{\mathbf{\Psi}}^{\phantom \dagger}}_{#1}}
\newcommand{\opbfPsidag}[1]{{\widehat{\mathbf{\Psi}}^{\dagger}}_{#1}}
\newcommand{\Hc}{\mathrm{H.c.}}
\newcommand{\Ham}{\widehat{H}}
\newcommand{\mathU}{\mathbb{U}}
\newcommand{\gs}{\mathrm{\scriptstyle gs}}
\newcommand{\ex}{\mathrm{\scriptstyle ex}}

\graphicspath{{./}{./Fig/}}

\begin{document}
\author{Ruiyi Wang}
\affiliation{SISSA, Via Bonomea 265, I-34136 Trieste, Italy}

\author{Vincenzo Roberto Arezzo}
\affiliation{SISSA, Via Bonomea 265, I-34136 Trieste, Italy}

\author{Kiran Thengil}
\affiliation{SISSA, Via Bonomea 265, I-34136 Trieste, Italy}

\author{Giovanni Pecci}
\affiliation{SISSA, Via Bonomea 265, I-34136 Trieste, Italy}

\author{Giuseppe E. Santoro}
\affiliation{SISSA, Via Bonomea 265, I-34136 Trieste, Italy}
\affiliation{CNR-IOM - Istituto Officina dei Materiali, Consiglio Nazionale delle Ricerche, c/o SISSA Via Bonomea 265, 34136 Trieste, Italy}
\affiliation{International Centre for Theoretical Physics (ICTP), P.O.Box 586, I-34014 Trieste, Italy}

\title{From Exponential to Quadratic: Optimal Control for a Frustrated Ising Ring Model}

\begin{abstract}
Exponentially small spectral gaps are known to be the crucial bottleneck for traditional Quantum Annealing (QA) based on interpolating between two Hamiltonians, a simple driving term and the complex problem to be solved, with a linear schedule in time. One of the simplest models showing exponentially small spectral gaps was introduced by Roberts {\em et al.}, PRA \textbf{101}, 042317 (2020): a ferromagnetic Ising ring with a single frustrating antiferromagnetic bond. A previous study of this model (C\^ot\'e {\em et al.}, QST \textbf{8}, 045033 (2023)) proposed a continuous-time diabatic QA, where optimized non-adiabatic annealing schedules provided good solutions, avoiding exponentially large annealing times. In our work, we move to a digital framework of Variational Quantum Algorithms, and present two main results:
1) we show that the model is {\em digitally controllable} with a scaling of resources that grows {\em quadratically} with the system size, achieving the exact solution using the Quantum Approximate Optimization Algorithm (QAOA);
2) We combine a technique of quantum control --- the Chopped RAndom Basis (CRAB) method --- and digitized quantum  annealing (dQA) to construct {\em smooth} digital schedules yielding optimal solutions with very high accuracy. 
\end{abstract}

\maketitle

\section{Introduction}
Optimization problems are of both theoretical interest and practical importance, as they lie at the core of many real-world challenges across diverse fields, including finance, engineering, machine learning, and healthcare. The complexity of systems and limited computational resources continually inspire advances in techniques to solve these problems. Quantum algorithms are expected to shed light on this field, and recent studies \cite{abbas2024challenges,blekos2024review,klug2023quantum} highlight their potential to improve solutions and efficiency in diverse problems, a prominent candidate being quantum annealing (QA)~\cite{finnila_quantum_1994,kadowaki1998quantum,Santoro_SCI02,Santoro_2006,Albash_RMP18}.

QA is a continuous-time quantum computational process designed to solve optimization problems. 
It is traditionally based on the adiabatic theorem, which governs the dynamics of the system under slow changes of a time-dependent Hamiltonian: by initializing the system in the ground state of a trivial Hamiltonian $\Ho_\mathrm{drive}$, it aims at adiabatically evolving the system into the ground state of a more complex Hamiltonian $\Ho_\mathrm{targ}$ encoding the solution to the optimization problem. 
The standard interpolating Hamiltonian is written as
\begin{equation}
     \Ho(t) = s(t) \, \Ho_\mathrm{targ} + (1-s(t)) \, \Ho_\mathrm{drive} \,,
\label{eqn:ann_hamiltonian_s}
\end{equation}
where $s(t)=t/\tau$ is the linear schedule with $s(0)=0$ and $s(\tau)=1$, and $\tau$ is the total annealing time.

The bottleneck of traditional QA appears when the spectral gap of the interpolating Hamiltonian decreases exponentially with the system size, leading to exponentially increasing annealing times ~\cite{Caneva_PRB2007,Knysh_NatComm16,Knysh_PRA2020,Zamponi_QA:review}. 
In the context of continuous-time Schr\"{o}dinger dynamics, improvements to conventional QA have been proposed.
A common strategy is to redesign annealing schedules $s(t)$, to improve the traditional linear schedule, by e.g., 
adopting a linear piecewise decomposition \cite{Matsuura_PRA2021,Cote_2023}, expanding over a polynomial basis \cite{Quiroz_PRA2019,Lucignano_PRA2022}, inserting pausing intervals within the schedule \cite{Passarelli_PRB2019} or via a Fourier decomposition \cite{Caneva_PRA2011,Montangero_dCRAB_PRA2015}.

Moving away from the continuous-time framework, 
two prominent methods are digitized Quantum Annealing (dQA) \cite{Nature_dQA,Mbeng_dQA_PRB2019} and the Quantum Approximate Optimization Algorithm (QAOA) \cite{farhi_quantum_2014}. 
Although conceptually distinct, they share the same alternating Hamiltonian {\em Ansatz} 
\begin{equation} \label{eqn:dQA_QAOA_ansatz}
|\Psi_{\Ptrot}(\btheta)\rangle =
\nep^{-i\theta^x_\Ptrot \Ho_x} 
\nep^{-i\theta^z_\Ptrot \Ho_z} \, 
\cdots 
\nep^{-i\theta^x_1 \Ho_x} 
\nep^{-i\theta^z_1 \Ho_z}  \, |\Psi_0\rangle \;,
\end{equation}
where the final state $|\Psi_{\Ptrot}(\btheta)\rangle$ is obtained by applying a depth-$2\Ptrot$ digital quantum circuit to the initial state $|\Psi_0\rangle$. 
Here, $\Ho_z=\Ho_\mathrm{targ}$ and $\Ho_x=\Ho_\mathrm{drive}$ are non-commuting Hamiltonians that involve, respectively, the $\PauliSigma^z$ and $\PauliSigma^x$ Pauli operators.
The final state depends on the $2\Ptrot$ parameters 
$\btheta=(\btheta^x,\btheta^z)$, where $\btheta^x=(\theta^x_1\cdots\theta^x_{\Ptrot})$ and 
$\btheta^z=(\theta^z_1\cdots\theta^z_{\Ptrot})$.

QAOA belongs to a wider class of Variational Quantum Algorithms (VQA)~\cite{cerezo_variational_2021} and was originally developed to solve combinatorial optimization problems. dQA ~\cite{Nature_dQA}, on the other hand, focuses on simulating the QA dynamics digitally.
The connection between dQA and QAOA has been investigated in Refs.~\cite{Mbeng_dQA_PRB2019,pecci2024beyond}. 

In both of these digitized approaches, the threshold for the state controllability of the model \cite{Dalessandro2007,MARGOLUS1998188} --- defined in this framework as the minimum number of unitary transformations required to reach the exact solution --- plays a fundamental role. It is essential to evaluate algorithm efficiency and to analyze resource scaling with system size \cite{Larocca2022diagnosingbarren}. 
For a generic model with an exponentially large Hilbert space, this quantity is very difficult to estimate. 
However, for specific models, symmetries can be exploited to determine the critical depth required to reach the target state \cite{mbeng2019optimal, niu2019optimizingqaoasuccessprobability}.

In this work, we present results, using the alternating {\em Ansatz} in Eq.~\eqref{eqn:dQA_QAOA_ansatz}, on the frustrated-ring model proposed in \cite{Knysh_PRA2020}, see also \cite{Cote_2023}, as the simplest problem exhibiting exponentially small spectral gaps as the number of spins increases. 
The model is a non-uniform Ising ring, where all the couplings except one are ferromagnetic, with a single antiferromagnetic coupling introducing frustration. 

We first use QAOA to characterize the controllability properties of this system in terms of the number of layers of the quantum circuit needed to exactly reach the desired ground state. 
We explicitly determine the critical depth $\Ptrot^{\mathrm{cr}}_1$ required to have a non-zero probability of reaching the target ground state, with a random initialization of the variational QAOA parameters $\btheta$. Remarkably, $\Ptrot^{\mathrm{cr}}_1$ scales quadratically with the number of spins, $\Ptrot^{\mathrm{cr}}_1=(\Nsites^2-1)/4$, completely bypassing the exponential bottleneck associated with the standard QA solution of the model. 

Next, we pursue the goal of finding smooth digital schedules for the parameters $\btheta$  in Eq.~\eqref{eqn:dQA_QAOA_ansatz}, which optimize the average energy $E_{\Ptrot}(\btheta) = \langle \Psi_{\Ptrot}(\btheta)| \Ho_z | \Psi_{\Ptrot}(\btheta)\rangle$.
A well-established quantum control technique, previously proposed and developed for continuous-time dynamics, is the Chopped RAndom Basis (CRAB) \cite{Caneva_PRA2011,Doria_PRL2011,Montangero_dCRAB_PRA2015,koch_quantum_2022}.
In the context of QA, CRAB is used to decompose the annealing schedules $s(t)$ into a finite set of smooth basis functions, such as Fourier modes, turning the functional minimization problem into a finite-dimensional one. 
We adopted this technique --- in its dressed CRAB (dCRAB) \cite{Montangero_dCRAB_PRA2015} flavor ---
in a digitized QA dynamics framework.
We show how combining CRAB with dQA produces smooth digital schedules 
$\btheta$ yielding results that outperform a randomly initialized QAOA in the regime $\Ptrot<\Ptrot^{\mathrm{cr}}_1$;
the approach provides high-accuracy results also in the controllable regime $\Ptrot>\Ptrot^{\mathrm{cr}}_1$. 
We then characterize the digital dynamics generated by such smooth solutions in terms of ground state populations and shortcut-to-adiabaticity, unveiling the non-adiabatic nature of the optimal strategy implemented by such dQA-CRAB approach. 
Finally, we introduce an effective annealing time for the smooth digital dQA-CRAB protocol and compare our results with diabatic quantum annealing in continuous time~\cite{Cote_2023}. We find that, for the same level of accuracy, our method exhibits better scaling with the system size.

The paper is organized as follows. 
Section \ref{sec:model} introduces the frustrated-ring model. 
Section \ref{sec:methodology} presents our methodology: the CRAB approach to digital schedule control in dQA (dubbed dQA-CRAB), \ref{sec:CRAB}, and QAOA from random starting initial points (dubbed QAOA-RAND), \ref{sec:QAOA}. 
Next, in Sec.~\ref{sec:results}, we present our results.
We start with our evidence of digital controllability through QAOA-RAND,
\ref{sec:controllability}, and then move to present the digitally smooth schedules produced by dQA-CRAB, \ref{sec:smooth_schedules}.
Next, we discuss the shortcut-to-adiabaticity ~\cite{STA_Review_RMP2019,TORRONTEGUI2013117} mechanism behind the dQA-CRAB protocol, \ref{sec:populations}, and its superiority with respect to the competing continuous-time diabatic QA approach of Ref.~\cite{Cote_2023}, \ref{sec:time}.
Finally, Section~\ref{sec:conclusions} summarizes our key findings and discusses future developments.

\section{The frustrated-ring model} \label{sec:model}
\begin{figure}[htp]
\includegraphics[width=0.7\columnwidth]{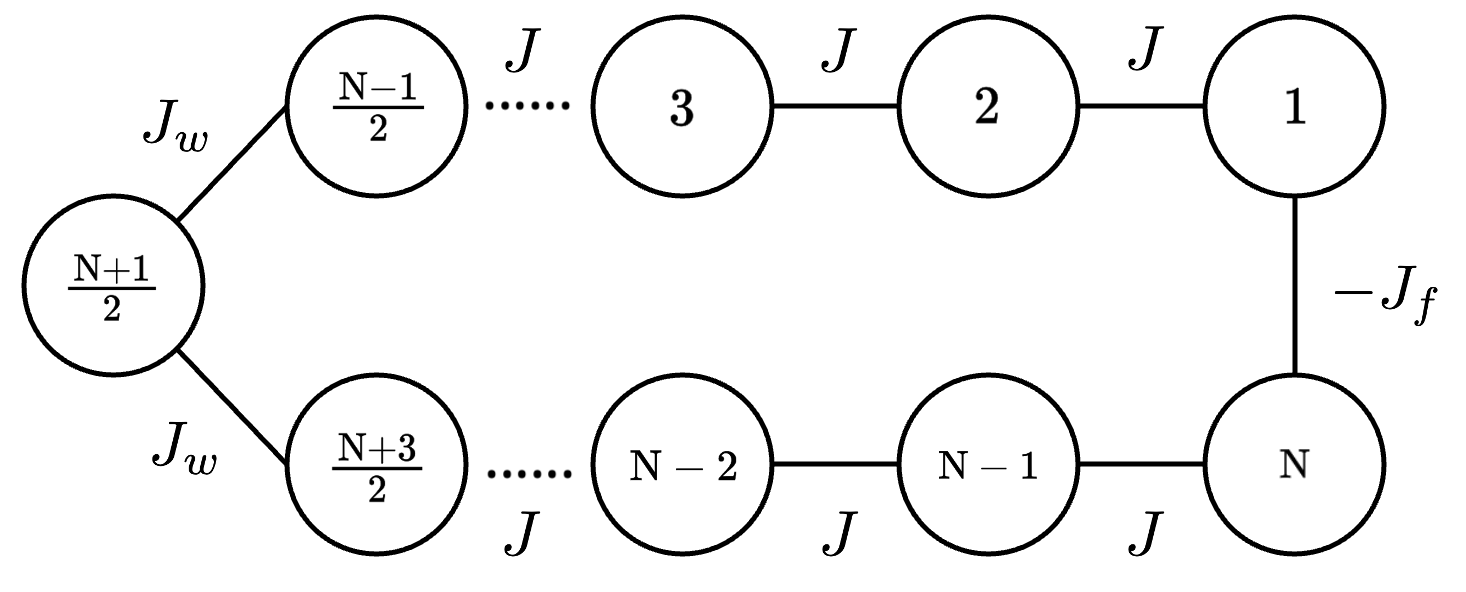}
\caption{A schematic diagram for the frustrated-ring model with an odd number of sites $\Nsites$. 
Most of the couplings between adjacent spins are uniform and ferromagnetic, $J$, except for the two central sites, coupled by weaker ferromagnetic bonds, $J_w$, and the first and last spins which are coupled by an antiferromagnetic (frustrating) bond $-J_f$.}
\label{fig_notes:model}
\end{figure}

Consider an \em odd \em number of spins $\Nsites$ arranged in a ring, as shown in Fig.~\ref{fig_notes:model}. 
This is the frustrated-ring model proposed in \cite{Knysh_PRA2020} and recently discussed in \cite{Cote_2023}. 
The problem Hamiltonian reads: 
\begin{equation}
    \Ho_\mathrm{targ} = \Ho_z = -\sum_{j=1}^\Nsites J_j \PauliSigma^z_j\PauliSigma^z_{j+1} \;.
\end{equation}
The couplings $J_j$ between adjacent spins are:
\begin{equation}
  J_{j} =
    \begin{cases}
      J_w & \text{if $j=(\Nsites\pm 1)/2$}\\
      -J_f & \text{if $j=\Nsites$}\\
      J & \text{otherwise}
    \end{cases}       \;.
\end{equation}
Here, $J$ is the ferromagnetic coupling between most sites, $J_w$ denotes the weaker ferromagnetic coupling between the two central sites at $j=(\Nsites\pm 1)/2$, and $J_f$ is the single frustrating (anti-ferromagnetic) coupling at site $\Nsites$. 
(The notation of Refs.~\cite{Knysh_PRA2020} and \cite{Cote_2023} have $J_f=J_R$ and $J_w=J_L$.)
We choose $0<J_f<J_w<J$, specifically requiring that $JJ_f>J_w^2$, which leads to the spin-glass bottleneck regime studied in Ref.~\cite{Knysh_PRA2020}. 
For the rest of this work we take, following Ref.~\cite{Cote_2023}, $J=1$, our unit of energy, $J_w=0.5$ and $J_f=0.45$.
The ground state is the ferromagnetic state with energy $E_{\gs}
=-(\Nsites-3)J-2J_w+J_f$; the first excited state has two domain walls, at a central site 
$j=(\Nsites+1)/2$ and at $j=N$, and is separated by an energy gap: $E_{\ex}-E_{\gs}\equiv\Delta_1=2(J_w-J_f)$. 

We write the driving term in the annealing Hamiltonian \eqref{eqn:ann_hamiltonian_s} as:
\begin{equation}
    \Ho_\mathrm{drive} = \Ho_x = -h \sum_{j=1}^\Nsites\PauliSigma^x_j \;.
\end{equation}

We now use the Jordan-Wigner transformation \cite{mbeng2024quantum} to transform $\Ho_z$ and $\Ho_x$ into quadratic fermionic Hamiltonians: 
\begin{equation} \label{eqn:Hzz_fermions}
\Ho_{z} = - \sum_{j=1}^{\Nsites} J_j 
\Big( \opcdag{j} \opc{j+1} \,+\, \opcdag{j} \opcdag{j+1}  \,+\, {\Hc} \Big) \;,
\end{equation}
with boundary condition $\opc{\Nsites+1}=(-1)^{\prm+1} \opc{1}$,
and 
\begin{equation}\label{eqn:Hx_fermions}
 \Ho_{x} = h \sum_{j=1}^{\Nsites} \Big( \opcdag{j} \opc{j} - \opc{j} \opcdag{j} \Big)  \;,
\end{equation}
where $\hat{c}^\dagger_j$ and $\hat{c}_j$ are fermionic creation and annihilation operators. Here, $\prm=0$ or $1$ denotes even or odd fermion parity sectors of the Hilbert space, $(-1)^{\prm}=\nep^{i\pi \Nfermions}$, where $\Nfermions=\sum_j \opcdag{j}\opc{j}$ is the total number of fermions. 
Notice that the ground state of $\Ho_{x}$ is the  fermionic vacuum $|0\rangle$ (which has $\prm=0$) for $h>0$, and the fully occupied fermionic state $|\mathrm{F}\rangle=\prod_j \opcdag{j}|0\rangle$ for $h<0$: this has $\prm=1$ when $\Nsites$ is odd.
In our work we will assume that $h<0$, setting it to $h=-J$, and restrict ourselves to the sub-sector with $\prm=1$, where the ground state of the interpolating Hamiltonian in Eq.~\eqref{eqn:ann_hamiltonian_s} stays and the whole dynamics takes place.

Using the Nambu formalism~\cite{mbeng2024quantum}, the fermionic Hamiltonians $\Ho_z$ and $\Ho_x$ are cast into matrix form:
\begin{eqnarray} \label{quadratic-H:eqn}
\Ho_{x/z} = \opbfPsidag{} \, \mathH_{x/z} \, \opbfPsi{} \;,
\end{eqnarray}
where $\mathH_{x/z}$ are $2\Nsites\times 2\Nsites$ Hermitian matrices and $\opbfPsi{} = (\opc{1}, \dots, \opc{\Nsites}, \opcdag{1}, \dots, \opcdag{\Nsites})^\transpose$, is a $2\Nsites$-dimensional column vector combining fermionic annihilation operators $\opc{1},\dots,\opc{\Nsites}$ and creation operators $\opcdag{1},\dots,\opcdag{\Nsites}$. 
For the transverse field part we have:
\begin{equation}
    \mathH_x = 
    \left( \begin{array}{cc} \A_x & \mathbf{0} \\
                           \mathbf{0} & -\A_x \end{array} \right) \;,
\end{equation}
where $\A_x$ is purely diagonal, here proportional to the identity since the transverse field is uniform:
\begin{equation}
    \A_x = h \mathbf{1} =-\mathbf{1} \;,
\end{equation}
where we set $h=-J=-1$.

For the target term, with $\prm=1$, we have:
\begin{equation}
    \mathH_z = 
    \left( \begin{array}{cc} \A_z & \B_z \\
                           -\B_z & -\A_z \end{array} \right) \;,
\end{equation}
where $\A_z$ and $\B_z$ are $\Nsites\times \Nsites$ real  symmetric or anti-symmetric matrices, respectively, and their only non-zero elements are given by:
\begin{equation} \label{AB_z}
\left\{
\begin{array}{l}
(\A_z)_{j,j+1} = (\A_z)_{j+1,j} = -\displaystyle
J_j/2 \vspace{4mm} \\
(\B_z)_{j,j+1} = -(\B_z)_{j+1,j} = - \displaystyle
J_j/2
\end{array}
\right. \;,
\end{equation}
with additional matrix elements, dependent on the fermionic parity, and given by:
\begin{equation}
(\A_z)_{\Nsites,1} = (\A_z)_{1,\Nsites} = (-1)^{\prm} J_{\Nsites}/2 
=+J_f/2 
\end{equation}
and
\begin{equation}
(\B_z)_{\Nsites,1} = -(\B_z)_{1,\Nsites} = (-1)^{\prm} J_{\Nsites}/2 
=+J_f/2 \;.
\end{equation}

\begin{figure}[htp]
\includegraphics[width=0.7\columnwidth]{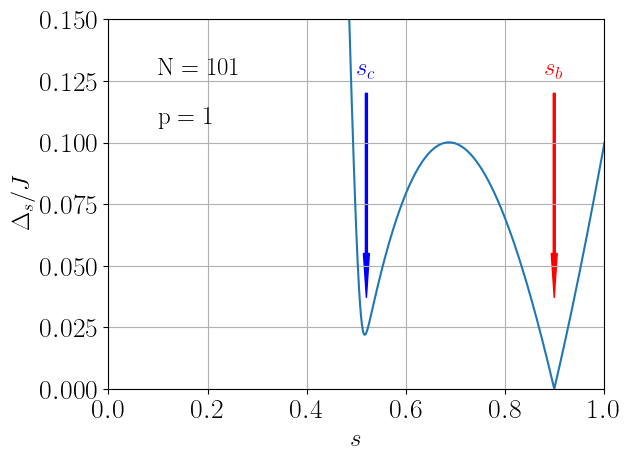}
\vspace{-5mm}
\caption{The spectral gap $\Delta_s$ between ground state and first excited state of a frustrated-ring model of size $\Nsites=101$ in the $\prm=1$ fermionic sub-sector, as a function of the interpolation parameter $s$. }
\label{fig:gap_NambuN101}
\end{figure}

Figure \ref{fig:gap_NambuN101} shows the spectral gap $\Delta_s$ between the ground state and 1st excited state of the interpolating Hamiltonian $s\Ho_z+(1-s)\Ho_x$ in the $\prm=1$ subsector, for the frustrated-ring model with $\Nsites=101$ sites (see Appendix~\ref{app:Nambuspectrum} for details).  
Notice that there are two small gaps. 
The first gap occurs at $s_c\approx 0.5$ and decays polynomially, as $1/\Nsites$: it is the critical gap separating the quantum paramagnetic phase ($s<s_c$) from the frustrated ferromagnetic phase ($s>s_c$). 
The second, at $s_b\approx 0.9$, is the exponentially small gap providing the spin-glass bottleneck discussed in Ref.~\cite{Knysh_PRA2020}. 
It is predicted~\cite{Knysh_PRA2020} to be located at:
\begin{equation}
    \label{eqn:position_mingap}
    s_b = \Bigg(1 + \frac{(J^2 - J_w^2)(J_w^2 - J_f^2)}{J J_f (J^2 + J_f^2 - 2J_w^2)}\Bigg)^{-1} \hspace{-1mm} \approx 0.89872068 \;,
\end{equation}
(the numerical value corresponds to $J_w/J=0.5$ and $J_f/J=0.45$), and to decay exponentially in $\Nsites$, as:
\begin{equation}
    \label{eqn:min_gap}
    \Delta_{s_b}  \mathop{\propto}_{\Nsites \to \infty}\left[ \frac{J(J_w^2-J_f^2)}{J_f(J^2-J_w^2)} \right]^\Nsites \;.
\end{equation}

\section{Methodology} \label{sec:methodology}
In continuous-time QA we have a time-dependent interpolating Hamiltonian
\begin{equation}
     \Ho(t) = s(t) \, \Ho_z + (1-s(t)) \, \Ho_x \;,
\label{eqn:ann_hamiltonian}
\end{equation}
where $s(0)=0$, $s(\tau)=1$, $\tau$ being the total annealing time. 
The system's state $|\Psi(t)\rangle$ follows a unitary Schr\"odinger dynamics 
$i\hbar \frac{\ud}{\ud t} |\Psi(t)\rangle = \Ho(t) |\Psi(t)\rangle$. 
An optimal control problem can be formulated as follows: We aim at an optimal schedule $s(t)$ such that the average energy 
\begin{equation}
E^{\fin}(\tau) = \langle \Psi(\tau) | \Ho_z | \Psi(\tau)\rangle 
\label{eqn:final_avg_energy}
\end{equation}
is minimal, at a given final time $\tau$. Since the quantum state $| \Psi(\tau)\rangle$ and hence the energy $E^{\fin}(\tau)$ depend on $s(t)$ --- a continuous function of $t$ --- this optimal control in continuous time is a {\em functional optimization problem}.  

The continuous-time approach by Côté {\em et al.} \cite{Cote_2023} parametrizes the annealing schedule $s(t)$ as a piecewise linear decomposition with a finite number of parameters. 
The parametrization starts from randomly selecting $\Ptrot-1$ values $s_p$, in the interval $[0,1)$, 
with $p=1\cdots\Ptrot-1$,
which are assigned to $\Ptrot-1$ equally spaced times $t_p=p \Delta_t=p \tau/\Ptrot$ in the time interval $[0,\tau]$: $s_p\equiv s(t_p)$.
The initial schedule $s(t)$ is then the piecewise linear function $s^{\mathrm{p-lin}}(t)$ obtained by connecting these $\Ptrot-1$ points $(t_p,s_p)$ with the two endpoints $(t_0=0,s(0)=0)$ and $(t_\Ptrot=\tau,s(\tau)=1)$. 
Different sets of random parameters $\{s_p\}$ are optimized with respect to the energy $E^{\fin}(\tau)$ of the output state $|\Psi(\tau)\rangle$, and the set that produces the lowest energy is selected as a candidate schedule $s^{\mathrm{p-lin}}(t)$. 
In subsequent iterations one doubles $\Ptrot \to 2\Ptrot$ by adding intermediate points in all linear segments of $s^{\mathrm{p-lin}}(t)$, and then optimizing the 
$2\Ptrot-1$ internal points $s_p$ to obtain a new 
piecewise-linear $s^{\mathrm{p-lin}}(t)$. 
In practice, starting from $\Ptrot=4$ and trying $N_{\mathrm{real}}=10$ realizations of 
$\{s_p\}$, one takes the best $s^{\mathrm{p-lin}}(t)$, and then moves to $2\Ptrot=8$ with $2\Ptrot-1=7$ internal points to be optimized; and so forth, until the final energy $E^{\fin}(\tau)$ associated with 
$s^{\mathrm{p-lin}}(t)$ differs from the actual ground state $E_{\gs}$ by less than a specific threshold $\Delta_E$.

\subsection{Digitized QA dynamics with CRAB} \label{sec:CRAB}

Let us now move towards a digitized QA dynamics~\cite{Nature_dQA,Mbeng_dQA_PRB2019,mbeng2019optimal,pecci2024beyond}. 
Suppose first that the annealing time $\tau$ is divided into a large number $\Ptrot$ of small times steps (not necessarily identical) $\Delta_p$, with $p=1\cdots\Ptrot$. 
The evolution operator associated to such time step is
approximately given by (setting $\hbar=1$):
\[
\nep^{-i\Delta_p (s_p\Ho_z+(1-s_p)\Ho_x)} \xrightarrow{\mathrm{Trotter}} 
\nep^{-i\theta^x_p \Ho_x} \nep^{-i\theta^z_p \Ho_z} \;,
\]
where $s_p=s(t_p)$ is the value of the schedule parameter at an appropriate point $t_p$ inside the time-interval (for instance, the middle point), and the $\theta^{x,z}_p$ are given, in a lowest order Trotter decomposition, as:
\begin{equation} \label{eqn:theta_Trotter}
    \theta^x_p = \Delta_p (1-s_p) \;,
    \hspace{10mm}
    \theta^z_p = \Delta_p s_p \;.
\end{equation}
Inverting these relationships we get:
\begin{equation} \label{eqn:schedule_inversion}
    \Delta_p=\theta^x_p+\theta^z_p \;,
    \hspace{10mm}
    s_p = \frac{\theta^z_p}{\theta^x_p+\theta^z_p} \;.
\end{equation}
The final state reached by applying the evolution operators for all time steps would then be given by:
\begin{equation} \label{eqn:dQA_state}
|\Psi_{\Ptrot}^{\dQA}\rangle =
\Uo(\theta^x_\Ptrot,\theta^z_\Ptrot) \, 
\cdots 
\Uo(\theta^x_1,\theta^z_1) \, |\Psi_0\rangle \;,
\end{equation}
where
\begin{equation} \label{eqn:1st_Trotter}
\Uo(\theta^x_p,\theta^z_p)=
\nep^{-i\theta^x_p \Ho_x} \nep^{-i\theta^z_p \Ho_z}  \;.
\end{equation}
Details on how such digital dynamics is implemented in the Bogoliubov-de Gennes formalism~\cite{mbeng2024quantum} for the frustrated-ring model after a Jordan-Wigner transformation are given in the Appendix~\ref{app:dynamics_Nambu}.

The standard linear schedule $s(t)=t/\tau$ in a digitized-QA setting would lead to 
\begin{equation} \label{eqn:theta_linear}
    \theta^{x,0}_p = \Delta_t \Big(1 - \textstyle\frac{t_p}{\tau}\Big) \;,
    \hspace{10mm}
    \theta^{z,0}_p = \Delta_t \textstyle\frac{t_p}{\tau} \;,
\end{equation}
where we set the time-grid $t_p = (p-\frac{1}{2})\Delta_t$, for $p=1,...,\Ptrot$, while $\tau=\Ptrot\Delta_t$.
By borrowing a well-established quantum control technique, the Chopped RAndom Basis (CRAB) \cite{Caneva_PRA2011,Doria_PRL2011,Montangero_dCRAB_PRA2015,koch_quantum_2022} previously proposed and developed in continuous-time dynamics we might write an improved {\em Ansatz} for the parameters by adding a finite number $\Nbasis$ of Fourier modes over the previous linear schedule 
\begin{equation} \label{eqn:CRAB_1}
\begin{array}{l}
\theta^{x,1}_p
= \rmC_0^{x} \theta^{x,0}_p 
+ \Big(1 - \frac{t_p}{\tau}\Big)
{\displaystyle\sum_{n=1}^{\Nbasis}} \rmC^{x}_n \sin(\omega^{\scriptscriptstyle(1)}_n t_p) 
\vspace{2mm} \\
\theta^{z,1}_p = \rmC_0^{z} \theta^{z,0}_p + \frac{t_p}{\tau}
{\displaystyle\sum_{n=1}^{\Nbasis}} \rmC^{z}_n \sin(\omega^{\scriptscriptstyle(1)}_n t_p)
\end{array}
\;,
\end{equation}
where $t_p/\tau=(p-\frac{1}{2})/\Ptrot$, and we incorporate $\Delta_t$ in the free coefficients $\rmC_0^x$ and $\rmC_0^z$.
The $2\Nbasis+2$ variational parameters $\C=(\rmC_0^x,\rmC_0^z,\C^x,\C^z)$, with $\C^{x/z}=(\rmC^{x/z}_1,\cdots,\rmC^{x/z}_{\Nbasis})$,  should be optimized to minimize the average energy
\begin{equation} \label{eqn:E_dQA}
E_{\Ptrot}(\C) = \langle \Psi_{\Ptrot}^{\dQA} | \Ho_z | \Psi_{\Ptrot}^{\dQA} \rangle \;.
\end{equation}
The choice of how large $\Nbasis$ should be and how to choose the $\Nbasis$ frequencies $\omega^{\scriptscriptstyle(1)}_n$ is a matter that we addressed by numerical experiments. 
We came out with a good recipe, which is to set $\Nbasis=\Ptrot$, and to generate $x=\omega^{\scriptscriptstyle(1)}_n\tau/\pi$ from a Gamma probability distribution 
\begin{align}
    \Gamma(x; \alpha, \beta) = 
\begin{cases} 
\frac{ x^{\alpha-1} e^{-x/\beta}}{\beta^\alpha\Gamma(\alpha)} & x \geq 0 \vspace{3mm}\\
0 & x < 0 
\end{cases} \;,
\end{align}
where $\alpha>0$ is the shape parameter, $\beta>0$ is the scale parameter, and
$\Gamma(\alpha)$ the Gamma function. 
We took $\alpha=3/2$ and $\beta=4$, so that the probability distribution is peaked at $x_{\max}=(\alpha-1)\beta=2$.

An important advantage of the digitized dynamics, over the continuous-time one, is that \textbf{gradients} of the energy $E_{\Ptrot}(\C)$ can be easily calculated analytically, see Appendix~\ref{app:gradients},
and given to the Broyden-Fletcher-Goldfarb-Shanno (BFGS) algorithm \cite{nocedal1999numerical} which we use to optimize the average energy. 

The CRAB procedure can be iteratively improved
by adopting a \textbf{dressed-CRAB} (dCRAB) approach~\cite{Montangero_dCRAB_PRA2015}, which adds new variational directions in the description of the control parameters: 
\begin{equation} \label{eqn:CRAB_2}
\begin{array}{l}
\theta^{x,2}_p
= \rmC_0^{x,2} \theta^{x,1}_p 
+ \Big(1 - \frac{t_p}{\tau}\Big)
{\displaystyle\sum_{n=1}^{\Nbasis}} \rmC^{x,2}_n \sin(\omega^{\scriptscriptstyle(2)}_n t_p) 
\vspace{2mm} \\
\theta^{z,2}_p = \rmC_0^{z,2} \theta^{z,1}_p + \frac{t_p}{\tau}
{\displaystyle\sum_{n=1}^{\Nbasis}} \rmC^{z,2}_n \sin(\omega^{\scriptscriptstyle(2)}_n t_p)
\end{array}
\;,
\end{equation}
with $(\btheta^{x,1},\btheta^{z,1})$ fixed from the previous CRAB optimization stage, and 
$2\Nbasis+2$ new variational parameters 
$\C=(\rmC_0^{x,2},\rmC_0^{z,2},\C^{x,2},\C^{z,2})$,
to be optimized so as to minimize 
$E_{\Ptrot}(\C)$ in Eq.~\eqref{eqn:E_dQA}.
The new fixed frequencies $\omega^{\scriptscriptstyle(2)}_n$ appearing in
Eq.~\eqref{eqn:CRAB_2}, it turns out, 
are crucial in adding larger frequency components to the control fields: in this second dCRAB iteration we generate 
$x=\omega^{\scriptscriptstyle(2)}_n\tau/\pi$ 
from $\Gamma(x,\alpha=\frac{3}{2},\beta=20)$, which
is peaked at $x_{\max}=10$. 

\begin{figure}[htp]
\includegraphics[width=0.7 \columnwidth]{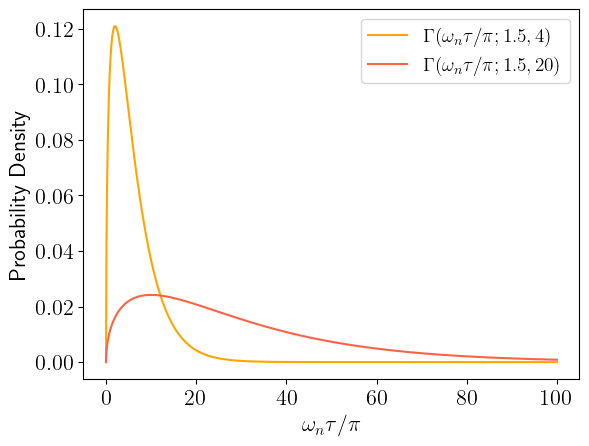}
\vspace{-5mm}
\caption{The two gamma distributions from which we generate the frequencies for the two dCRAB iterations.}
\label{fig:gammas}
\end{figure}

Figure \ref{fig:gammas} shows the two Gamma distributions from which we generate the frequencies for the two dCRAB iterations: $\Gamma(x,\frac{3}{2},4)$ for $\omega^{\scriptscriptstyle(1)}_n\tau/\pi$ in Eq.~\eqref{eqn:CRAB_1}, 
and
$\Gamma(x,\frac{3}{2},20)$ for $\omega^{\scriptscriptstyle(2)}_n\tau/\pi$
in Eq.~\eqref{eqn:CRAB_2}.

The algorithm described above will be referred to as dQA-CRAB.

\subsection{QAOA} \label{sec:QAOA}
The Quantum Approximate Optimization Algorithm (QAOA) \cite{Farhi_arXiv2014} is a variational quantum algorithm designed to solve combinatorial optimization problems by approximating the ground state of the problem Hamiltonian. It prepares the final quantum state using a parametrized sequence of $2\Ptrot$ alternating quantum gates which is identical to that appearing in
Eq.~\eqref{eqn:dQA_state}, i.e.,
\begin{equation} \label{eqn:QAOA_state}
|\Psi_{\Ptrot}^{\QAOA}(\btheta)\rangle =
\Uo(\theta^x_\Ptrot,\theta^z_\Ptrot) \, 
\cdots 
\Uo(\theta^x_1,\theta^z_1) \, |\Psi_0\rangle \;,
\end{equation}
where
\begin{equation} \label{eqn:1st_Trotter_bis}
\Uo(\theta^x_p,\theta^z_p)=\nep^{-i\theta^x_p \Ho_x} 
\nep^{-i\theta^z_p \Ho_z}  \;,
\end{equation}
with $p=1,\cdots,\Ptrot$. 
The conceptual difference with respect to Eq.~\eqref{eqn:dQA_state} is that here the 
$\btheta=(\btheta^x,\btheta^z)$, where 
$\btheta^x=(\theta^x_1\cdots\theta^x_{\Ptrot})$ and 
$\btheta^z=(\theta^z_1\cdots\theta^z_{\Ptrot})$ are $\Ptrot$-dimensional vectors, are supposed to be {\em free} variational parameters  with respect to which we should minimize the average energy:
\begin{equation} \label{eqn:E_QAOA}
E_{\Ptrot}^{\QAOA}(\btheta) = \langle \Psi_{\Ptrot}^{\QAOA}(\btheta)| \Ho_z | \Psi_{\Ptrot}^{\QAOA}(\btheta)\rangle  \;.
\end{equation}
The minimization of the cost function is, in general, non-trivial due to the complexity of the $2\Ptrot$-dimensional search space. 
Many strategies have been discussed in the literature: for a brief discussion of them, see Ref.~\cite{pecci2024beyond}.
Here, we present results obtained on the frustrated-ring model by initializing the optimization process with {\em randomly} generated starting points, employing the BFGS algorithm \cite{nocedal1999numerical} with analytically calculated gradients to achieve efficient convergence toward a local or global minimum. 
We will call this algorithm QAOA-RAND.

\section{Results} \label{sec:results}
To assess the quality of the solutions obtained, we will use, 
as a figure of merit, a rescaled residual energy:
\begin{equation} \label{eqn:res_energy}
\epsilon^{\res}_{\Ptrot} 
= \frac{E_{\Ptrot} - E_{\gs}}{\Nsites} \;,
\end{equation}
where $E_{\Ptrot}$ is either the 
dQA-CRAB, Eq.~\eqref{eqn:E_dQA}, or 
QAOA-RAND, Eq.~\eqref{eqn:E_QAOA}, 
optimal solution,
$E_{\gs}=-(N-3)J-2J_w+J_f$ is the ground state energy of the frustrated-ring model, and $\Nsites$ is the number of sites. 

\subsection{Digital controllability through QAOA-RAND} \label{sec:controllability}

\begin{figure*}[htp]
\centering
\subfloat[$\Ptrot<\Ptrot^{\mathrm{cr}}_1$]{\includegraphics[width=0.45\linewidth]{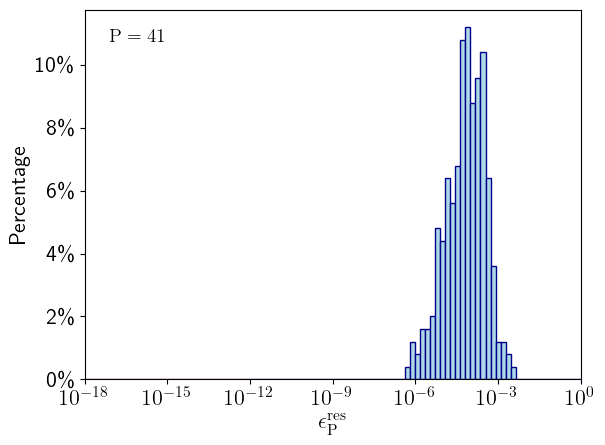}}
\quad
\subfloat[$\Ptrot=\Ptrot^{\mathrm{cr}}_1$]{\includegraphics[width=0.442\linewidth]{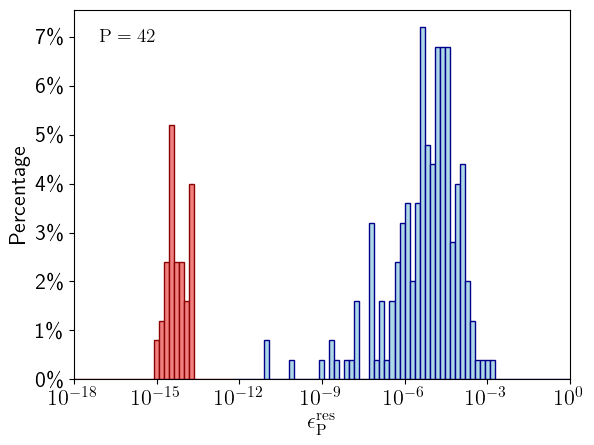}}
\quad
\subfloat[$\Ptrot=\Ptrot^{\mathrm{cr}}_2$]{\includegraphics[width=0.45\linewidth]{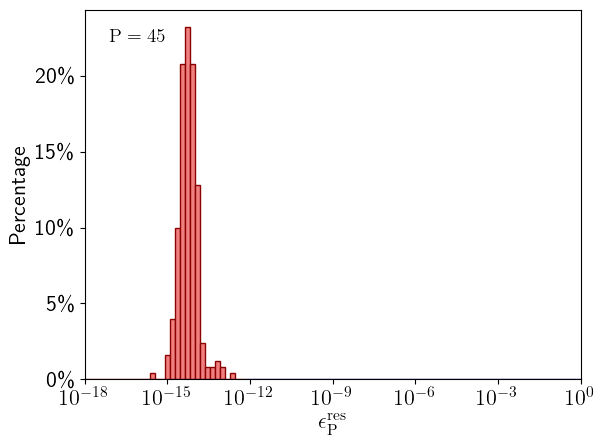}}
\quad
\subfloat[$\Ptrot^{\mathrm{cr}}_1$ vs $\Nsites^2$]{\includegraphics[width=0.435\linewidth]{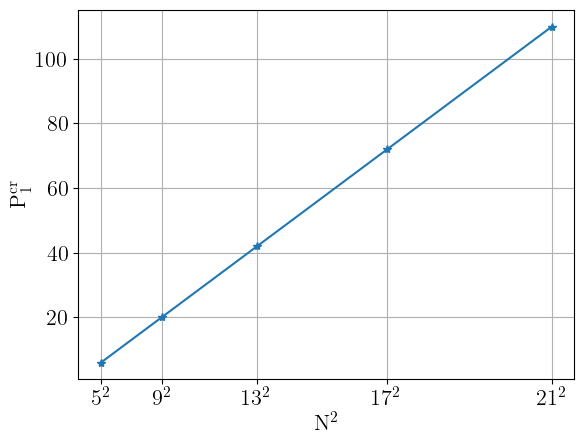}}
\caption{(a) - (c) Histograms displaying $250$ residual energies $\epsilon_\Ptrot^\res$ obtained by QAOA-RAND. Here we consider a frustrated-ring model with $\Nsites=13$ and $\Ptrot=41,42,45$. 
Red columns correspond to residual energies lower than $10^{-12}$, which we take as numerical zero.
(d) $\Ptrot^\mathrm{cr}_1$ scales quadratically with system size $\Nsites$.}
\label{fig_notes:hist}
\end{figure*}

We start presenting evidence of a digital controllability of the ground state of the model, by employing QAOA-RAND.
Figure (\ref{fig_notes:hist}) (a) - (c) shows histograms of the residual energy for a ring of $\Nsites=13$ and $\Ptrot$ from $41$ to $45$: each histogram comprises 250 QAOA optimization from random starting points (QAOA-RAND).
We observe that $\Ptrot=42$ is the smallest $\Ptrot$ in which a (numerically) zero residual energy, that is, the exact ground state, could be reached. 
We denote as $\Ptrot^{\mathrm{cr}}_1$ this critical $\Ptrot$ at which full control of the system's state is achieved.
For $\Nsites=13$, we find that $\Ptrot^{\mathrm{cr}}_1=42$. 
As $\Ptrot$ increases, the percentage of simulations achieving zero residual energies increases. 
At $\Ptrot=45$, any random starting parameters in QAOA would ensure that the exact ground state is reached. 
We call $\Ptrot^{\mathrm{cr}}_2$ the value of $\Ptrot$ at which the QAOA landscape appears to be simple enough that any random initial point leads to the solution. 

The same pattern appears for larger system sizes $\Nsites$. 
Numerically, we find that $\Ptrot^{\mathrm{cr}}_1$
grows \textbf{quadratically} with the system size $\Nsites$, in stark contrast with the exponentially closing gap $\Delta_{s_b}$, which implies exponentially long annealing times for a linear schedule QA. 
As numerically shown in Figure (\ref{fig_notes:hist})(d) (see final discussion for a rationalization of this behavior) the expression for $\Ptrot^{\mathrm{cr}}_1$ in terms of $\Nsites$ appears to be: 
\begin{equation}
\label{eqn:P_cr_1}
\Ptrot^{\mathrm{cr}}_1=\frac{\Nsites^2-1}{4} \;.
\end{equation}
Concerning $\Ptrot^\mathrm{cr}_2$, its dependence on $\Nsites$ is less clear,
partly due to the non-sharp definition of the threshold we consider for a ``numerical zero''. 

\subsection{dQA-CRAB smooth optimal schedules}
\label{sec:smooth_schedules}
In analogy with the Trotter-discretized expressions in 
Eq.~\eqref{eqn:schedule_inversion}, we define, for a general digitized algorithm, even away from the Trotter limit, 
the total evolution time $\tau$ in terms of $\Delta_p$ as: 
\begin{equation} \label{eqn:tau_sp_def}
    \tau=\sum_{p=1}^{\Ptrot} \Delta_p = \sum_{p=1}^{\Ptrot} (\theta^x_p+\theta^z_p) \;,
\end{equation}
where $p=1,\dots,\Ptrot$ labels the sequence of time steps.

\begin{figure}[htp]
\includegraphics[width=0.7\columnwidth]{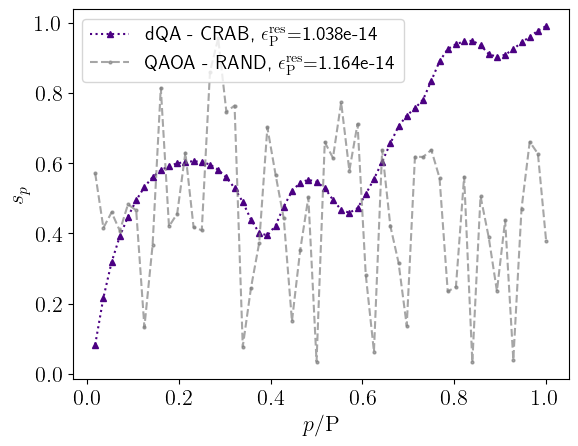}
\vspace{-3mm}
\caption{Comparison of smooth and random schedules obtained from dQA-CRAB and QAOA-RAND, respectively. Here we consider the frustrated-ring model for $\Nsites=13$ and $\Ptrot=56>\Ptrot^{\mathrm{cr}}_1$. 
}
\label{fig_notes:s_compare}
\end{figure}

We show in Fig.~\ref{fig_notes:s_compare} the schedules 
$\{s_p\}$ defined in Eq.~\ref{eqn:schedule_inversion} and produced using QAOA-RAND and dQA-CRAB on the frustrated-ring model with $\Nsites=13$ and $\Ptrot=56$.  
As expected dQA-CRAB produces smooth schedules, while QAOA-RAND is agnostic to any regularity.
Observe that the dQA-CRAB schedule $s_p$ has an 'up-down-up' shape qualitatively similar to the $s^{\mathrm{p-lin}}(t)$ found in Ref.~\cite{Cote_2023} (see their Figure 6). 

\begin{figure}[htp]
\includegraphics[width=0.7\columnwidth]{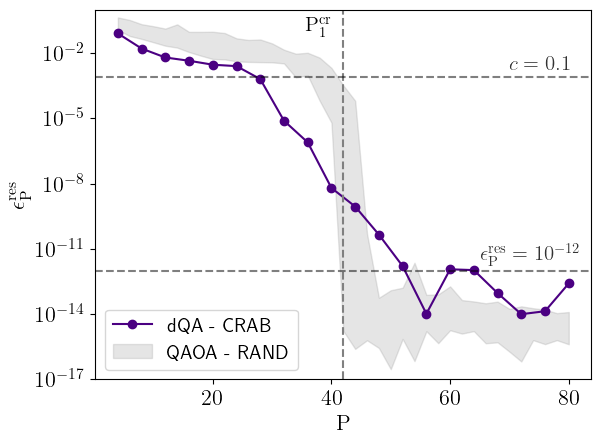}
\vspace{-3mm}
\caption{Comparisons of residual energy $\epsilon^{\res}_{\Ptrot}$ versus $\Ptrot$ for the two methods: QAOA-RAND and dQA-CRAB. Here $\Nsites=13$. The dashed vertical line indicates the critical depth $\Ptrot^{\mathrm{cr}}_1=42.$} 
\label{fig_notes:N13_eres_P}
\end{figure}

To better understand the quality of the smooth solutions obtained with dQA-CRAB, we also show in Fig.~\ref{fig_notes:N13_eres_P} a direct comparison of the residual energies $\epsilon_{\Ptrot}^{\res}$ obtained from the two methods, QAOA-RAND and dQA-CRAB, for fixed $\Nsites=13$. 
The shaded region for the QAOA-RAND data spans from the minimum to the maximum of the residual energy values obtained from $100$ different random realizations for each $\Ptrot$. 
Notice how before $\Ptrot^{\mathrm{cr}}_1$, dQA-CRAB yields solutions notably better than even the best case of randomly initialized QAOA. 
This implies that for small $\Ptrot$ 
we achieve reasonably good solutions with a smooth schedule produced from dQA-CRAB,
while QAOA-RAND should be avoided, as it leads to poor-quality local minima. 
When $\Ptrot$ increases and crosses $\Ptrot^{\mathrm{cr}}_1$,
dQA-CRAB shows a less abrupt decrease in the residual energies obtained, compared to the steep drop obtained in QAOA-RAND. At and beyond $\Ptrot=56$, dQA-CRAB reaches residual energies around or below $10^{-12}$, as shown by the lower horizontal dashed line.

Figure~\ref{fig_notes:N13_eres_P} also shows, as the upper horizontal dashed line labeled $c=0.1$, the performance criterion used by C\^ot\'e {\em et al.} in \cite{Cote_2023} (their equation (9)): 
\begin{equation}
\label{eqn:criteria_c}
E_{\Ptrot} - E_{\min}
\leq c \, 
\Delta_{s=1} \;,
\end{equation}
where $\Delta_{s=1}=2(J_w-J_f)$ is the energy gap between the ground state and the first excited state of the problem Hamiltonian \cite{Cote_2023}.
When $c=0.1$, $\Nsites=13$ and $J_w-J_f=0.05$, such threshold corresponds to 
$\epsilon^{\res}_{\Ptrot}\leq 7.6723\times 10^{-4}$, 
and is reached by dQA-CRAB at $\Ptrot=28$. 
A more detailed comparison of the efficiency of our algorithm against the continuous-time strategy of Ref.~\cite{Cote_2023} will be discussed in Sec.~\ref{sec:time}.

\begin{figure*}[htp]
\centering
\centering
\subfloat[$\Nsites=13$, $\Ptrot=40$]{\includegraphics[width=0.45\linewidth]{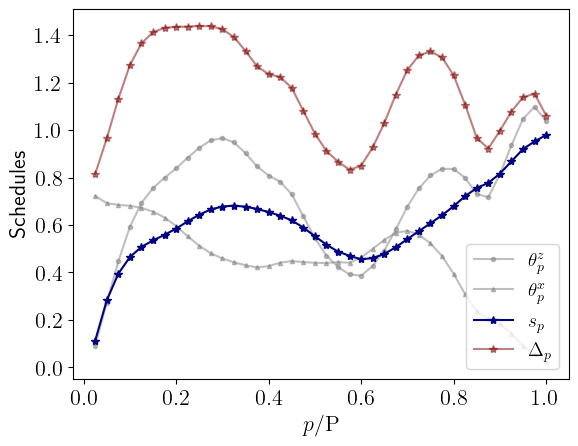}}
\quad
\subfloat[$\Nsites=45$, $\Ptrot=100$]{\includegraphics[width=0.45\linewidth]{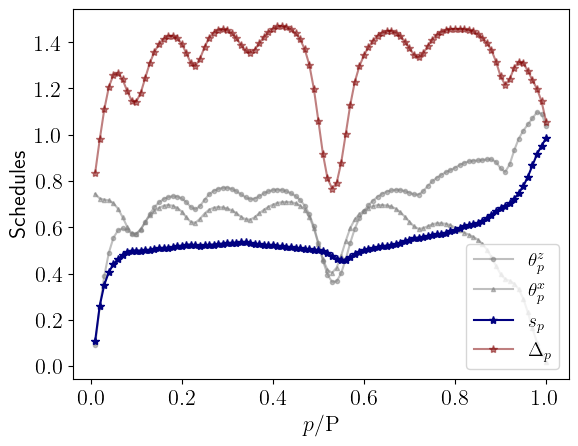}}
\quad
\subfloat[$\Nsites=13$, $\Ptrot=40$]{\includegraphics[width=0.45\linewidth]{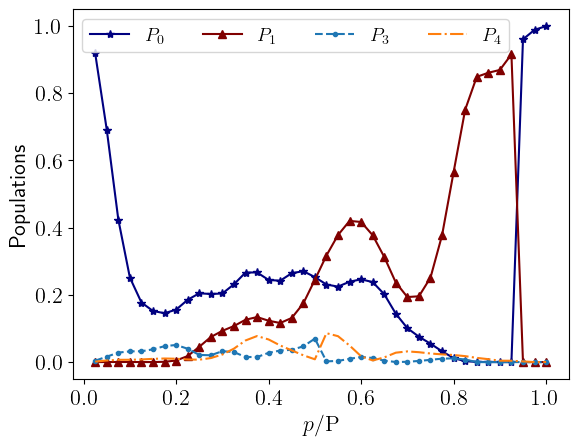}}
\quad
\subfloat[$\Nsites=45$, $\Ptrot=100$]{\includegraphics[width=0.45\linewidth]{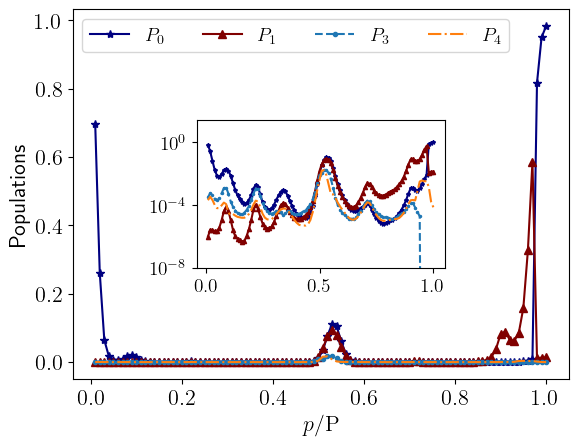}}
\quad
\subfloat[$\Nsites=13$]{\includegraphics[width=0.45\linewidth]{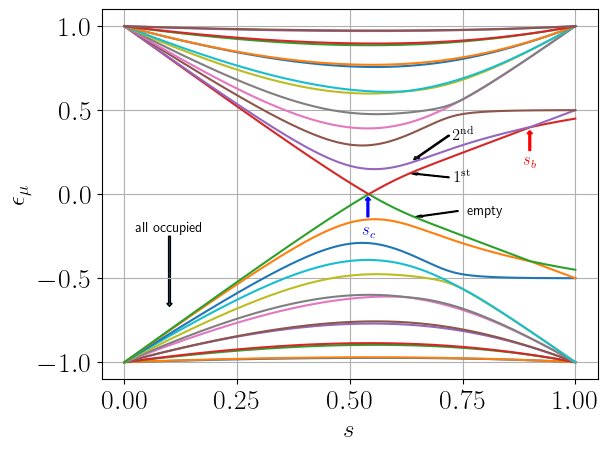}}
\quad
\subfloat[$\Nsites=45$]{\includegraphics[width=0.45\linewidth]{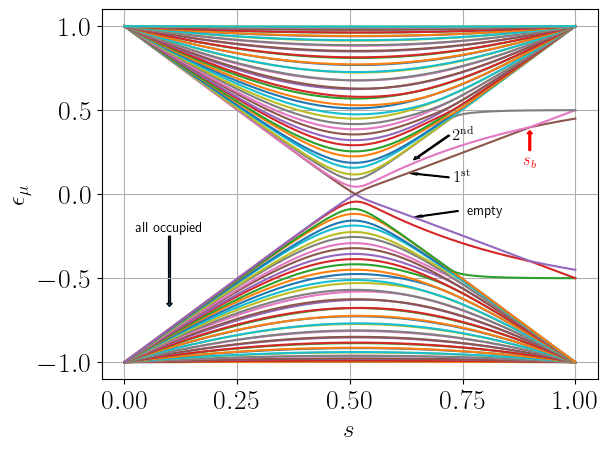}}
\caption{(a,b) Optimal dQA-CRAB parameters, (c,d) populations of the instantaneous ground ($P_0$) and lowest exited states ($P_1\cdots P_4$), (e,f) Bogoliubov-de Gennes eigenvalue spectrum for the interpolating Hamiltonian $s\Ho_z+(1-s)\Ho_x$ versus $s$. 
The optimal schedules in (a,b) are generated using dQA-CRAB: two super-iterations of dressed-CRAB, where frequencies are sampled from gamma distributions, as described in Sec.~\ref{sec:CRAB}.
The residual energy for the case $\Nsites=45$, $\Ptrot=100$ is  $\epsilon^{\res}_{\Ptrot}= 1.2\times 10^{-4}$.}
\label{fig_notes:pops}
\end{figure*}

\subsection{Populations of the instantaneous eigestates in dQA-CRAB}
\label{sec:populations}

To study what kind of dynamics, adiabatic or otherwise, the dQA-CRAB control parameters $(\theta^x_p,\theta^z_p)$ generate at each discrete ``time step'' $p$, we would like to access the digitized equivalent of the {\em instantaneous} eigenstates of the evolving Hamiltonian. 
Unfortunately, at each step $p$ we apply two unitary operators, 
$\nep^{-i\theta^x_p \Ho_x} \nep^{-i\theta^z_p \Ho_z}$, 
and this complicates the unambiguous identification of such ``instantaneous eigenstates''. 
To bypass such difficulty, we consider, given the control parameters $(\theta^x_p,\theta^z_p)$, an effective Hamiltonian $\Ho_p^\mathrm{eff}$ defined by the first-order term of the Baker-Campbell-Hausdorff (BCH) formula, i.e.:
\begin{equation}
    \Ho_p^\mathrm{eff} = \frac{1}{\Delta_p}\left(\theta^x_p\Ho_x +\theta^z_p\Ho_z\right) =
    s_p\Ho_z + (1-s_p)\Ho_x\;,
\end{equation}
where $s_p$ and $\Delta_p$ are defined in Eq.~\eqref{eqn:schedule_inversion}. 
Let $\ket{\phi^{j,\eff}_p}$ be the $j$th instantaneous eigenstate of the effective Hamiltonian $\Ho_p^\mathrm{eff}$.
We then define the population of the instantaneous eigenstates $\ket{\phi^{j,\eff}_p}$ as:
\begin{equation}
   P_j(p) = |\langle \phi^{j,\eff}_p | \Psi_{p} \rangle |^2 \;,
   \label{eqn:populations}
\end{equation}
where $|\Psi_p\rangle=\Uo(\theta^x_p,\theta^z_p) \cdots \Uo(\theta^x_1,\theta^z_1)|\Psi_0\rangle$.
The eigenstates $\ket{\phi^{j,\eff}_p}$ are constructed, for small systems, by performing exact diagonalization; 
for larger systems, we rely on the Jordan-Wigner mapping followed by a Bogoliubov transformation, and employ the techniques explained in Ref.~\cite{mbeng2024quantum}[Appendix C] to calculate overlaps between states. 

Figure~\ref{fig_notes:pops}(a,b) show the optimal dQA-CRAB schedules ($\theta^x_p$, $\theta^z_p$, and associated $s_p$ and $\Delta_p$) for $\Nsites=13$ and $\Ptrot=40$ (a), and for $\Nsites=45$ and $\Ptrot=100$ (b).  
Figure~\ref{fig_notes:pops}(c,d) show the corresponding 
discrete-time ground ($P_0$) and low excited states ($P_1\cdots P_4$) instantaneous populations, for the same two values of $\Nsites$ and $\Ptrot$.
An adiabatic algorithm would keep a large ground state population $P_0$ throughout the evolution. 
In contrast, we see that our dQA-CRAB dynamics implements, effectively, a shortcut-to-adiabaticity (STA)~\cite{STA_Review_RMP2019,TORRONTEGUI2013117} mechanism: the ground state population $P_0$ rapidly decreases, hence promoting excitations to higher levels, to increase sharply at the final stages of the dynamics, where the closing spectral gap is encountered. 

To illustrate such a phenomenon in more detail, we refer the reader to the Bogoliubov-de Gennes spectrum shown in Fig.~\ref{fig_notes:pops}(e,f), again for the two values of $\Nsites$. 
We see that there is a clear singularity in the spectrum at $s_c\approx 0.5$, with a \textbf{vanishing gap} in the Nambu spectrum around the particle-hole symmetry point $E=0$.
This implies a true level crossing, which we will comment on shortly. 
For $s<s_c$, the system's state is obtained by occuping all the $\Nsites$ negative energy Bogoliubov-de Gennes eigenvalues, with a large gap to the first exited state. 
However, as mentioned, at $s_c$ a level crossing implies that the occupied state with energy $-\epsilon_1$ for $s<s_c$ is {\em adiabatically transformed} into the state at energy $+\epsilon_1$ when $s>s_c$, hence it remains {\em occupied}. 
Symmetrically, the (empty) state with energy $+\epsilon_1$ for $s<s_c$, is adiabatically transformed into the state at energy $-\epsilon_1$ for $s>s_c$, and remains empty.
Such an adiabatic transformation is quite detrimental for the quality of the evolution: the occupied state at
energy $+\epsilon_1$ when $s>s_c$ suffers from the closing gap bottleneck at $s_b$, where it would almost unavoidably become excited. 
The optimal strategy is that the system {\em preempts} such occurrence by exciting the system, around $s\approx s_c$, ideally to the $2^{\mathrm{nd}}$ level (see Fig.~\ref{fig_notes:pops})(e,f): this will directly lead, when the system reaches the avoided crossing bottleneck at $s_b$, towards the final ground state. 
The populations of eigenstates calculated for the case with $\Nsites=13$, see Fig.~\ref{fig_notes:pops}(a), follow quite closely such a two-state STA scenario, as witnessed by the small involvement of higher excited states. 
What is more surprising is the result for $\Nsites=45$, see Fig.~\ref{fig_notes:pops}(b), where the vicinity of higher exited states in the Nambu spectrum at $s_c$, see Fig.~\ref{fig_notes:pops}(f), leads to appreciable populations of quite a number of excited states: indeed, both $P_0$ and $P_1$ remain quite small everywhere, with a final surge of $P_1$ right before the location of the bottleneck $s_b$, where the standard two-state STA scenario apparently takes place. 

\subsection{Annealing time comparison} \label{sec:time}
\begin{figure}[htp]
    \centering
\includegraphics[width=0.7\columnwidth, trim={0 0 0 0}, clip]{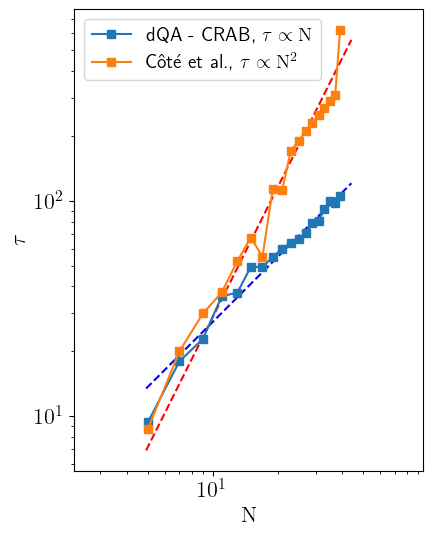}
\caption{Comparison of the annealing time $\tau$ needed to reach the threshold $c=0.1$, see Eq.~\eqref{eqn:criteria_c}, for our dQA-CRAB method and the method of C\^ot\'e {\em et al.}~\cite{Cote_2023} (data taken from their Fig. 5), versus the system size $\Nsites$.}
\label{fig_notes:time}
\end{figure}

In the digitized setting, the total annealing time is defined, see Eq.~\eqref{eqn:tau_sp_def}, as:
\begin{equation} \label{eqn:annealing_time}
    \tau = \sum_{p=1}^{\Ptrot} 
    \left(\theta^x_p+ \theta_p^z\right) \;.
\end{equation}
We compare the efficiency of our dQA-CRAB algorithm against the continuous-time approach by C\^ot\'e {\em et al.}~\cite{Cote_2023} in terms of the annealing time required to reach the threshold $c=0.1$ as defined in \ref{eqn:criteria_c}. 
Figure \ref{fig_notes:time} shows such comparison.  
While a quadratic scaling $\tau\propto\Nsites^2$ is observed for the continuous-time approach of Ref.~\cite{Cote_2023}, we see that our digital algorithm using the smooth schedules obtained from dQA-CRAB shows a linear scaling $\tau\propto \Nsites$. 

\section{Conclusions and perspectives} \label{sec:conclusions}

We have explored digitized QA approaches and QAOA to overcome the challenges posed by exponentially small spectral gaps, a common bottleneck to adiabatic evolutions in traditional QA~\cite{Zamponi_QA:review}. 
We have focused on the frustrated-ring model, proposed as the simplest class of Ising spin systems whose minimum gap decreases exponentially with the system size $\Nsites$. 
By exploiting the Jordan-Wigner mapping, we were able to perform an exact simulation of the digital dynamics up to quite large values of $\Nsites$.

As a first result, we leveraged QAOA to provide a detailed characterization of the different controllability ~\cite{Dalessandro2007} regimes of the model.
At any system size $\Nsites$, we identify two critical depths $\Ptrot^{\mathrm{cr}}_1$ and $\Ptrot^{\mathrm{cr}}_2>\Ptrot^{\mathrm{cr}}_1$ for the associated QAOA quantum circuit, defining three distinct regimes. 
For depths with $\Ptrot<\Ptrot^{\mathrm{cr}}_1$, QAOA with a random initialization of parameters $(\theta^x_p,\theta^z_p)$ never achieves the exact ground state of the model. 
As the depth of the circuit equals $\Ptrot^{\mathrm{cr}}_1$, the probability of reaching the exact ground state with a random initialization of QAOA starts to be finite, and continues to grow until the second critical value $\Ptrot^{\mathrm{cr}}_2$ is reached. 
Beyond this point, any random initialization of the angles $(\theta^x_p,\theta^z_p)$ leads to an optimal protocol that converges to the exact ground state. 
Furthermore, we numerically determined an exact expression for $\Ptrot^{\mathrm{cr}}_1=(\Nsites^2-1)/4$, which scales as the square of the number of spins. 
This offers a significant advantage over the exponentially long evolution time required in standard QA. 
A formal proof for the value of the critical depth $\Ptrot^{\mathrm{cr}}_1$ in the frustrated-ring model will be presented in Ref.~\cite{Arezzo_inpreparation}.
It suffices here to notice that the Bogoliubov-de Gennes theory behind the exact Jordan-Wigner mapping of this and related quantum Ising chain models predicts that the dynamics (both in discrete and continuous time) is fully encoded into a $2\Nsites \times 2\Nsites$ unitary transformation $\mathU$
: hence, a scaling with $\Nsites^2$ is quite to be expected. The precise value $\Ptrot^{\mathrm{cr}}_1=(\Nsites^2-1)/4$ follows from the additional reflection symmetry that the frustrated-ring model Hamiltonian possesses~\cite{Arezzo_inpreparation}. 

On the other hand, precisely determining the scaling of $\Ptrot^{\mathrm{cr}}_2$ results to be more complex. 
When the circuit depth reaches $\Ptrot^{\mathrm{cr}}_2$, every local minima of the average energy becomes degenerate with the global minimum, leading to a much less complex optimization landscape. Characterizing this process is challenging because, unlike $\Ptrot^{\mathrm{cr}}_1$, it is not strictly related to either the number of parameters required to describe the dynamics or to state controllability \cite{Dalessandro2007, ALBERTINI2002213}. Furthermore, any numerical claim about the scaling of $\Ptrot^{\mathrm{cr}}_2$ depends on the definition of numerical zero residual energy --- we set it to be $10^{-12}$ --- which is inherently arbitrary.  In this regard, having an analytical insight is essential to provide any scaling law for $\Ptrot^{\mathrm{cr}}_2$. 

The second result of our work concerns the design of optimal smooth digital schedules $(\theta^x_p,\theta^z_p)$. 
To do so, we employed dressed-CRAB~\cite{Montangero_dCRAB_PRA2015} quantum control techniques in our digitized QA approach. 
Beyond ensuring smoothness, this procedure, which we dubbed dQA-CRAB, outperforms a randomly-initialized QAOA for circuit depths below the critical threshold, $\Ptrot<\Ptrot^{\mathrm{cr}}_1$. 
In the controllable regime $\Ptrot>\Ptrot^{\mathrm{cr}}_1$, the smooth schedules produced with dQA-CRAB continue to perform very well, leading to final states very close to the actual ground state. 

In order to characterize the dynamics implemented by smooth dQA-CRAB solutions, we focused on the populations of the instantaneous eigenstates of the effective Hamiltonian associated with the digital dynamics. 
Unlike in adiabatic protocols, we see clear signs of a shortcut-to-adiabaticity mechanism taking place in our protocol.
Far before the closing of the spectral gap, the dQA-CRAB optimal schedules smoothly populate the excited states of the effective Hamiltonian spectrum. 
Subsequently, the system non-adiabatically crosses the small spectral gap, 
leading to population inversion and to a large occupation of the final ground state. 

Finally, we compared our dQA-CRAB with the continuous-time diabatic annealing algorithm proposed in \cite{Cote_2023}. 
We show that, for a fixed predefined accuracy threshold, our method requires a total annealing time $\tau$ that scales linearly with the system-size, $\tau\propto \Nsites$, as opposed to quadratically, $\tau\propto \Nsites^2$, as in \cite{Cote_2023}.

One of the most important messages of our study is that the spectral gap bottleneck, a nightmare of traditional QA approaches based on linear schedules \cite{Zamponi_QA:review}, has nothing to do with the actual computational difficulty of the problem at hand or to how hard is the controllability issue: the frustrated-ring Ising model is an example of a polynomially controllable problem with a severe exponentially small spectral gap bottleneck.

An interesting avenue to pursue is that of a \emph{quantum speed limit}
\cite{MARGOLUS1998188, giovannetti_qsl, levitin_qsl,del_campo_qsl,PhysRevLett.120.070402} in a digitized framework, i.e., the minimum circuit depth required to exactly reach a target state, starting from a fixed initial configuration. 
This quantity represents the ultimate limit for optimal digitized control protocols \cite{Caneva_PRL2009}. 
Previous work on specific models are known~\cite{mbeng2019optimal,niu2019optimizingqaoasuccessprobability,Wauters_PRA2020,Christandl_PRA2005}, but a more systematic understanding is mandatory. 

A detailed study of the optimal continuous-time evolution, and its associated quantum speed limit, would be an intriguing extension of this work, offering insights into the interplay between controllability in the digital setting and the fundamental limitations imposed by quantum speed limits.

Another promising avenue involves employing counterdiabatic driving~\cite{KOLODRUBETZ20171} into the current framework, see also \cite{Wurtz2022counterdiabaticity}, which could potentially further improve the accuracy of the solution and computational efficiency. Exploring this integration in both continuous and digitized settings may reveal novel ways to enhance the performance of the proposed quantum algorithms.

\begin{acknowledgments}
We acknowledge useful discussions with Glen B. Mbeng.
G.E.S. and K.T. acknowledge financial support from PNRR MUR project PE0000023-NQSTI. G.E.S. and G.P. acknowledge financial support from PRIN 2022H77XB7 of the Italian Ministry of University and Research, and from the QuantERA II Programme STAQS project that 
has received funding from the European Union’s H2020 research and innovation programme under Grant Agreement No 101017733.
G.E.S. acknowledges that his research has been conducted within the framework of the Trieste Institute for Theoretical Quantum Technologies (TQT).
\end{acknowledgments}

\appendix
\section{Low-energy spectrum in the Nambu formalism}
\label{app:Nambuspectrum}
We consider here only the odd-parity case, $\prm=1$.
We denote by $\epsilon_{\mu}>0$ the positive energy Bogoliubov-de Gennes eigenvalues~\cite{mbeng2024quantum}, ordered from the smallest, $\epsilon_1$, to the largest. 
Recall that the Bogoliubov-de Gennes Hamiltonian has the diagonal form~\cite{mbeng2024quantum}:
\begin{equation}
\Ho = \sum_{\mu=1}^{\Nsites} \epsilon_{\mu} \Big( \opgammadag{\mu} \opgamma{\mu} - \opgamma{\mu} \opgammadag{\mu} \Big) \;. 
\end{equation}
Referring to Fig.~\ref{fig_notes:pops}(e,f), we see that there is a clear singularity in the spectrum at $s_c\approx 0.5$, with a vanishing gap in the Nambu spectrum around the particle-hole symmetry point $E=0$, which implies a true level crossing, where $\epsilon_1(s=s_c)\equiv 0$.
Hence, the (occupied) state with energy $-\epsilon_1$ for $s<s_c$ is transformed into the state at energy $+\epsilon_1>0$ when $s>s_c$, hence it remains occupied. 
Symmetrically, the (empty) state with energy $+\epsilon_1$ for $s<s_c$, is transformed into the state at energy $-\epsilon_1$ for $s>s_c$, hence it remains empty.

For $s<s_c$ the ground state is the Bogoliubov vacuum obtained by occupying all the negative energy states:
\begin{equation} \label{eqn:gs_smalls}
    |\Psi_\gs\rangle = |\emptyset\rangle = \Big( \prod_{\mu=1}^{\Nsites} \opgamma{\mu} \Big) |0\rangle \;,
    \hspace{5mm} \Longrightarrow \hspace{5mm}
    E_{\gs}(s<s_c) = -\sum_{\mu=1}^{\Nsites} \epsilon_{\mu} \;.
\end{equation}
The first excited state (conserving the fermionic parity) is obtained by creating {\em two} fermions in the lowest two positive energy states, hence:
\begin{equation} \label{eqn:ex_smalls}
    |\Psi_{\ex}^{(21)}\rangle = \opgammadag{2} \opgammadag{1} |\emptyset\rangle \;,
    \hspace{5mm} \Longrightarrow \hspace{5mm}
     E_{\ex}^{(21)}(s<s_c) = -\sum_{\mu=1}^{\Nsites} \epsilon_{\mu} + 2(\epsilon_{2} + \epsilon_{1}) \;.
\end{equation}
Higher excited states are obtained similarly.

For $s>s_c$ the ground state is obtained by ``occupying'' the first positive eigenvalue. 
Formally, this amounts to performing a particle-hole transformation for the first level 
\begin{equation} \label{eqn:ph-gamma}
    \opgammatilde{1} = \opgammadag{1} \;, \hspace{10mm} \mbox{while} \hspace{10mm}
    \opgammatilde{\mu} = \opgamma{\mu} \hspace{10mm} \mbox{for} \; \mu=2,\cdots,\Nsites \;,
\end{equation}
with energies transformed accordingly as 
$\tilde{\epsilon}_1=-\epsilon_1<0$ and 
$\tilde{\epsilon}_{\mu}=\epsilon_{\mu}$ for $\mu=2,\cdots,\Nsites$.
The ground state $|\tilde{\emptyset}\rangle$ is annihilated by the $\opgammatilde{\mu}$ and the expressions in Eq.~\eqref{eqn:gs_smalls}\eqref{eqn:ex_smalls} can be directly used. 
For instance, the ground state energy is:
\begin{equation}
    E_{\gs}(s>s_c) = 
    -\sum_{\mu=1}^{\Nsites} \tilde{\epsilon}_{\mu} = 
    - \sum_{\mu=2}^{\Nsites} \epsilon_{\mu}
    + \epsilon_{1} \;.
\end{equation}
The first excited state is:
\begin{equation}
|\Psi_{\ex}^{(21)}\rangle = 
\opgammatildedag{2} \opgammatildedag{1} |\tilde{\emptyset}\rangle \;,
\end{equation}
with energy:
\begin{equation}
    E_{\ex}^{(21)}(s>s_c) =
    -\sum_{\mu=1}^{\Nsites} \tilde{\epsilon}_{\mu} + 2(\tilde{\epsilon}_{1} + \tilde{\epsilon}_{2})
    = -\sum_{\mu=2}^{\Nsites} \epsilon_{\mu} - \epsilon_{1} + 2 \epsilon_{2} \;.
\end{equation}
Hence, the gap between the excited and ground state is:
\begin{equation}
    \Delta_s^{(21)} = 
    \left\{ \begin{array}{ll} 2(\epsilon_{2}+\epsilon_{1}) & \hspace{10mm} \mbox{for}\; s<s_c \vspace{4mm} \\
    2(\epsilon_{2} -\epsilon_{1}) & \hspace{10mm} \mbox{for}\; s>s_c
    \end{array}
    \right. \;.
\end{equation}
This is the spectral gap shown in Fig.~\ref{fig:gap_NambuN101} for $\Nsites=101$.

\section{Digitized dynamics in the Nambu formalism}
\label{app:dynamics_Nambu}
We derive here the expression for the 
digitized dynamics of the transverse field random Ising model,
as implemented by the time-evolution operator:
\begin{equation} \label{Effective_ham_def}
\hat{U}_p^{\step} = 
\nep^{-i \theta_p^x \Ho_{x}}
\nep^{-i \theta_p^z \Ho_{z} }
\;, 
\end{equation}

The idea is that we move to the Heisenberg representation for the fermionic operators, in a way that closely follows the derivation of the time-dependent Bogoliubov-de Gennes (BdG) equations in Ref.~\cite{mbeng2024quantum}.
Recall that for a system represented by a quadratic fermionic Hamiltonian $\Ho(t)= \opbfPsidag{} \, \mathH(t) \, \opbfPsi{}$, the Heisenberg operators for the fermions can be written as:
\begin{equation} \label{eqn:BdG_tdep_1}
\opbfPsi{\Heis}(t) \equiv \left( \begin{array}{c} 
\opc{1\Heis}(t) \\ \vdots \\ \opc{ \scriptscriptstyle{\Nsites}\Heis}(t) \\ \opcdag{1\Heis}(t) \\ \vdots \\ \opcdag{\scriptscriptstyle{\Nsites}\Heis}(t)
\end{array}
\right) = \mathU(t)
\left( \begin{array}{c} \opgamma{1} \\ \vdots \\ \opgamma{\scriptscriptstyle{\Nsites}} \\ \opgammadag{1} \\  \vdots \\ \opgammadag{\scriptscriptstyle{\Nsites}}
\end{array}
\right) \;,
\end{equation}
where the Bogoliubov operators $\opgamma{\mu}$ annihilate the initial state $\opgamma{\mu}|\Psi_0\rangle=0$ and the $2\Nsites\times 2\Nsites$ unitary matrix $\mathU(t)$ satisfies the time-dependent BdG equations:
\begin{equation} \label{eqn:BdG_tdep_2}
i\hbar \frac{\ud}{\ud t} \mathU(t) = 2\, \mathH(t) \, \mathU(t) \;.
\end{equation}

From now on, we assume that we perform our digitized dynamics in the \textbf{odd fermion parity sector}, starting from the fully occupied fermionic state 
\[ 
|\Psi_0\rangle=|\mathrm{F}\rangle = \prod_j \opcdag{j} |0\rangle
\]
corresponding to the ground state of $\Ho_{x}$ in Eq.~\eqref{eqn:Hx_fermions} for $h=-J$.
The Bogoliubov-de Gennes operators diagonalizing the initial Hamiltonian $\Ho_x$ are then given by $\opgamma{j}=\opcdag{j}$. 
With the standard \cite{mbeng2024quantum} writing:
\[
\opbfPsi{} \equiv \left( \begin{array}{c} \opc{1} \\ \vdots \\ \opc{\Nsites} \\ \opcdag{1} \\ \vdots \\ \opcdag{\Nsites}
\end{array}
\right) = \mathU_0 
\left( \begin{array}{c} \opgamma{1} \\ \vdots \\ \opgamma{\Nsites} \\ \opgammadag{1} \\ \vdots \\ \opgammadag{\Nsites}
\end{array}
\right) \;,
\]
we deduce that the initial Bogoliubov transformation matrix appearing in Eq.~\eqref{eqn:BdG_tdep_1} is:
\begin{equation}
    \mathU_0 = 
    \left( \begin{array}{cc} \mathbf{0} & \mathbf{1} \\
                           \mathbf{1} & \mathbf{0} \end{array}
                           \right) \;.
\end{equation}

The unitary dynamics is such that at step $p=1\cdots \Ptrot$ we have a state:
\begin{equation}
    |\Psi_p\rangle = \hat{U}^{\step}_p |\Psi_{p-1}\rangle \;,
\end{equation}
with the initial condition $|\Psi_0\rangle$, and $\hat{U}^{\step}_p$ given by:
\begin{equation} 
\hat{U}^{\step}_p = 
\nep^{-i \theta_p^x \Ho_{x}}
\nep^{-i \theta_p^z \Ho_{z}} \;. 
\end{equation}
Such a dynamics can be seen as a series of {\em quantum quenches}, hence a special case of a general time-dependent $\Ham(t)$, for which the time-dependent Bogoliubov-de Gennes (BdG) equations ~\eqref{eqn:BdG_tdep_2} apply ~\cite{mbeng2024quantum}.
Consider now the single application of $\nep^{-i\theta \Ham}$
--- with a {\em time-independent} $\Ham$, either $\Ho_{z}$ or $\Ho_{x}$ --- 
to any Bogoliubov vacumm (Gaussian) state $|\Psi_0\rangle$. 
You can view it as the integration of a Schr\"odinger equation
\begin{equation}  \label{Schr:eq}
i \frac{d}{d\theta} |\Psi(\theta)\rangle = \Ham |\Psi(\theta)\rangle \;,
\end{equation}
with initial condition $|\Psi(\theta=0)\rangle$.
The corresponding BdG equation reads:
\begin{equation} \label{bog_theta:eqn}
i \frac{d}{d\theta} \mathU(\theta)
= 2 \, \mathH \,
\mathU(\theta),
\end{equation}
with boundary condition $\mathU(\theta=0)$
set by $|\Psi(\theta=0)\rangle$. Therefore, the explicit solution of the BdG equations reads:
\begin{equation}
    \mathU(\theta) = \nep^{-2i \theta \, \mathH} \,
     \mathU(\theta=0) \;.
\end{equation}
Let now $\mathU$ to be the unitary matrix formed by the diagonalization of $\mathH$:
\begin{equation}
\mathH = \mathU \, {\mathbb E}_{\diag} \, 
\mathU^{\dagger} \;.
\end{equation}
Then, it is simple to show that 
\begin{equation}
    \mathU(\theta) = \mathU \,
    \nep^{-2i \theta \, {\mathbb E}_{\diag}}
    \mathU^{\dagger} \, \mathU(\theta=0) \;.
\end{equation}

The successive application of the two operators comprising $\hat{U}^{\step}_p$ is therefore obtained 
from the diagonalizations, 
$\mathH_z=\mathU_z \, {\mathbb E}^z_{\diag} \, \mathU^{\dagger}_z$, 
and
$\mathH_x=\mathU_x \, {\mathbb E}^x_{\diag} \, 
\mathU^{\dagger}_x$. 
The latter is particularly simple, as the Hamiltonian is diagonal in the original fermions,
\[
{\mathbb E}^x_{\diag} = \mathH_x =
\left( \begin{array}{rr} -\mathbf{1} & \mathbf{0} \\
                           \mathbf{0} & \mathbf{1} \end{array}
                           \right) \;.
\]
Hence $\mathbb{U}_x={\mathbb 1}$, and therefore: 
\begin{equation}
\nep^{-2i \theta^x_p \mathH_x} =
\left( \begin{array}{cc} \nep^{+2i\theta^x_p } \identity  & \mathbf{0} \\ \mathbf{0} & \nep^{-2i\theta^x_p} \identity  \end{array} \right) \;.
\end{equation}
The diagonalization of $\Ham_{z}$ in the $\prm=1$ sector can be done by hand. 
Observe that in the original fermions, we can rewrite Eq.~\eqref{eqn:Hzz_fermions} as:
\begin{equation}
    \Ho_{z} = -\sum_{j=1}^{\Nsites} J_j
    \Big(\opcdag{j}-\opc{j}\Big) \Big(\opcdag{j+1}+\opc{j+1}\Big) 
    \;,
\end{equation}
where $\opc{N+1}=\opc{1}$.
Next, observe that the BdG combinations 
\begin{equation} \label{eqn:BdG_gammatilde}
\left\{ \begin{array}{rcl}
    \opgammatilde{j} &=& \frac{1}{2} \Big( \opcdag{j+1}+\opcdag{j}+\opc{j+1}-\opc{j}\Big)  \vspace{3mm} \\
    \opgammatildedag{j} &=& \frac{1}{2} \Big( \opcdag{j+1}-\opcdag{j}+\opc{j+1}+\opc{j} \Big)
    \end{array}
\right. \;,
\end{equation}
are such that 
\begin{equation}
\left\{ \begin{array}{rcl}
    \opgammatildedag{j}+\opgammatilde{j} &=&  \opcdag{j+1}+\opc{j+1} 
    \vspace{3mm} \\
    \opgammatildedag{j}-\opgammatilde{j} &=& \opc{j}-\opcdag{j}
    \end{array}
\right. \;.
\end{equation}
Hence, we can write:
\begin{align}
    \Ho_{z} = \sum_{j=1}^{\Nsites} J_j
    \Big( \opgammatildedag{j}-\opgammatilde{j} \Big) \Big( \opgammatildedag{j}+\opgammatilde{j} \Big)
    &= \sum_{j=1}^{\Nsites} J_j \Big( \opgammatildedag{j}\opgammatilde{j} - \opgammatilde{j}\opgammatildedag{j}\Big) 
    \label{eqn:Htarg_gamma}
    = 2\sum_{j=1}^{\Nsites} J_j \opgammatildedag{j}\opgammatilde{j} - \sum_{j=1}^{\Nsites} J_j \;.
\end{align}
Since $J_{\Nsites}=-J_f<0$, the ground state
energy is correct: $E_{\gs}=-\sum_{j=1}^{\Nsites} J_j=-(\Nsites-3)J-2J_w+J_f$. These $\opgammatilde{j}$ are precisely the particle-hole transformed operators invoked in Eq.~\eqref{eqn:ph-gamma}, and the ground state $|\tilde{\emptyset}\rangle$ of $\Ho_z$ is their vacuum state, $\opgammatilde{j}|\tilde{\emptyset}\rangle=0$.  
The BdG eigenvectors are conveniently organized as follows
~\cite{mbeng2024quantum}.
If you define the $\Nsites\times \Nsites$ blocks $\U_z$ and $\V_z$ of the $2\Nsites\times 2\Nsites$ BdG matrix $\mathU_z$:
\begin{equation}
\mathU_z = 
    \left( \begin{array}{cc} \U_z  & \V_z^* \\ \V_z & \U_z^*  \end{array} \right) \;,
\end{equation}
then, Eq.~\eqref{eqn:BdG_gammatilde} tells us that:
\begin{equation}
\U_z = \left( 
\begin{array}{cccccc}
-\onehalf  & 0 & 0 & \cdots & 0 & \onehalf  \\ 
\onehalf & -\onehalf & 0 & \cdots & \vdots & 0 \\ 
0 & \onehalf & -\onehalf & \cdots & \vdots & 0 \\ 
0  & 0 & \onehalf & \cdots & \vdots & 0  \\ 
\vdots & \vdots & \vdots & \cdots & \vdots & \vdots \\
0 & 0 & 0 & \cdots & -\onehalf & 0 \\
0 & 0 & 0 & \cdots & \onehalf & -\onehalf 
\end{array}
\right) 
\hspace{10mm} \mbox{and} \hspace{10mm}
\V_z = \left( 
\begin{array}{cccccc}
\onehalf & 0 & 0 & \cdots & 0 & \onehalf \\
\onehalf & \onehalf & 0 & \cdots & \vdots & 0  \\
0 & \onehalf & \onehalf & \cdots & \vdots & 0  \\
0 & 0 & \onehalf & \cdots & \vdots & 0 \\
\vdots & \vdots & \vdots & \cdots & \vdots & \vdots  \\
0 & 0 & 0 & \cdots & \onehalf & 0  \\
0 & 0 & 0 & \cdots & \onehalf & \onehalf 
\end{array}
\right) \;.
\end{equation}
The application of 
$ \hat{U}^{\step}_p = 
\nep^{-i \theta_p^x \Ho_{x}}
\nep^{-i \theta_p^z \Ho_{z}} $
is therefore carried out by:
\begin{equation} \label{eqn:Ustep}
\mathU^{\step}_{p} = 
\nep^{-2i \theta^x_p \mathH_x}
    \mathU_z \,
    \nep^{-2i \theta^z_p {\mathbb E}^z_{\diag}}
    \mathU^{\dagger}_z \;.
\end{equation}
By carrying out the algebra in Eq.~\eqref{eqn:Ustep}, the final block form of $\mathU^{\step}_{p}$ is found to be:
\begin{equation}
\mathU^{\step}_p = 
    \left( \begin{array}{cc} \U(\theta^x_p,\theta^z_p)  & \V^*(\theta^x_p,\theta^z_p) \\ \V(\theta^x_p,\theta^z_p) & \U^*(\theta^x_p,\theta^z_p)  \end{array} \right) \;,
\end{equation}
with:
\begin{align}
    \U(\theta^x_p,\theta^z_p) 
    &= \nep^{+2i\theta^x_p }  \Big(
    \U_z \bfD_p 
    \U_z^{\dagger}
     + \V_z^* \bfD^*_p
     \V_z^{\transpose} \Big) \nonumber \vspace{4mm} \\
     \V(\theta^x_p,\theta^z_p) 
     &= \nep^{-2i\theta^x_p}  \Big(
    \V_z \bfD_p 
    \U_z^{\dagger}
     + \U_z^* \bfD^*_p 
     \V_z^{\transpose} \Big)
     \;,
\end{align}
and:
\begin{align} \label{eqn:D_p}
\bfD_p(\theta^z_p) = 
\nep^{-2i\theta^z_p \diag(\epsilon^z_{\mu})} 
= 
\left( 
\begin{array}{cccc}
\nep^{-2i\theta^z_p J_1} & \cdots & 0 & 0 \\
\vdots  & \ddots & \vdots & \vdots  \\
0 & \cdots & \nep^{-2i\theta^z_p J_{\scriptscriptstyle{\Nsites-1}} } & 0  \\
0  & \cdots & 0 & \nep^{-2i\theta^z_p J_{\Nsites}}
\end{array}
\right) \;,
\end{align}
where we should keep in mind that $J_{\Nsites}=-J_f$, while all the other couplings are positive. 

The final solution for the digitized dynamics can be summarized as follows:
\begin{equation}
    \mathU_{\Ptrot} = 
    \mathU^{\step}_{\Ptrot} \mathU^{\step}_{\Ptrot-1} \cdots
    \mathU^{\step}_{2} \mathU^{\step}_{1} \mathU_{0} \;,
\end{equation}
where 
\begin{equation}
    \mathU_0 = 
    \left( \begin{array}{cc} \mathbf{0} & \mathbf{1} \\
                           \mathbf{1} & \mathbf{0} \end{array}
                           \right) \;,
\end{equation}
as set by the initial state of the computation. 

\section{Residual energy}
\label{app:residual}
We can quantify the quality of the solution by the residual energy:
\begin{equation}
    \epsilon^{\res}_p = \frac{1}{N} \Big( \langle \Psi_p | \Ham_{z} | \Psi_p \rangle - E_{\gs} \Big), 
\end{equation}
where $E_{\gs}$ is the target ground state energy, which for the frustrated-ring model is given by 
\[ 
E_{\gs}
=-\sum_{j=1}^{\Nsites} J_j 
=-\sum_{j=1}^{\Nsites-1} J_j + J_f \;, 
\]
with $J_{\Nsites}=-J_f$. 
Hence, we can write:
\begin{equation} \label{eqn:res_energy_2}
    \epsilon^{\res}_p = \frac{1}{\Nsites} \sum_{j=1}^{\Nsites} J_j 
    \langle \Psi_p | \big( 1-\PauliSigma^z_{j}\PauliSigma^z_{j+1} \big) | \Psi_p \rangle \;. 
\end{equation}
In terms of Jordan-Wigner fermions we have that:
\begin{equation}
    \langle \Psi_p | \PauliSigma^z_{j}\PauliSigma^z_{j+1} | \Psi_p \rangle \to 
    \langle \Psi_p | \big(\opcdag{j}\opc{j+1}+\opcdag{j}\opcdag{j+1} + \Hc \big) | \Psi_p \rangle \;.
    \nonumber
\end{equation}
One can define the elementary Green's functions:
\begin{equation}
    (\G_p)_{j,j'} = \langle \Psi_p | \opc{j} \opcdag{j'} | \Psi_p \rangle \;, 
    \hspace{10mm} 
     (\F_p)_{j,j'} = \langle \Psi_p | \opc{j} \opc{j'} | \Psi_p \rangle \;.
\end{equation}
From the t-dependent BdG equations~\cite{mbeng2024quantum}, we can express both Green's functions in terms of
the blocks of the matrix 
\[ 
\mathU_p= \mathU^{\step}_{p} \mathU^{\step}_{p-1} \cdots
\mathU^{\step}_{1} \mathU_{0} = 
 \left( \begin{array}{cc} \U_p  & \V_p^* \\ \V_p & \U_p^*  \end{array} \right) \;.
\]
Indeed:
\begin{equation}
    \G_p = \U_p \U_p^{\dagger} \;, \hspace{10mm}
    \F_p = \U_p \V_p^{\dagger} \;. 
\end{equation}
Hence, we can easily calculate 
\begin{equation}
    -\langle \Psi_p | \PauliSigma^z_{j}\PauliSigma^z_{j+1} | \Psi_p \rangle \to  \Big( (\F_p+\G_p)_{j,j+1} + \text{c.c.} \Big) \;.
\end{equation}
From this expression, the residual energy 
follows by appropriate sums over $j$, as implied by Eq.~\eqref{eqn:res_energy_2}.\\

\section{Gradients of the residual energy}
\label{app:gradients}
We now consider the residual energy $\epsilon^{\res}_{\Ptrot}$,
which is a function of the $2\Ptrot$ variational parameters $(\btheta^x,\btheta^z)$. 
We would like to calculate the $2\Ptrot$ gradients of such a quantity, so as to use gradient-based optimization algorithms. 
If $\theta^{\alpha}_p$, with $\alpha=x,z$ denotes the generic variable, we need to calculate:

\begin{widetext}
\begin{align}
-\frac{\partial}{\partial \theta^{\alpha}_p} 
\langle \Psi_{\Ptrot} | \PauliSigma^z_{j}\PauliSigma^z_{j+1} | \Psi_{\Ptrot} \rangle &\to  
\frac{\partial}{\partial \theta^{\alpha}_p} 
\Big( \U_{\Ptrot} (\U_{\Ptrot}^{\dagger} + \V_{\Ptrot}^{\dagger}) \Big)_{j,j+1} + \text{c.c.} \nonumber \\
&= \Bigg(  \Big(\frac{\partial}{\partial \theta^{\alpha}_p} \U_{\Ptrot} \Big) (\U_{\Ptrot}^{\dagger} + \V_{\Ptrot}^{\dagger})
+ \U_{\Ptrot} \frac{\partial}{\partial \theta^{\alpha}_p} (\U_{\Ptrot}^{\dagger} + \V_{\Ptrot}^{\dagger})
\Bigg)_{j,j+1} + \text{c.c.}
\;.
\end{align}
\end{widetext}

The relevant derivative of $\U_{\Ptrot}$ and $\V_{\Ptrot}$ 
are obtained from the corresponding blocks of
\begin{equation}
    \frac{\partial}{\partial \theta^{\alpha}_p} \mathU_{\Ptrot} =
    \mathU^{\step}_{\Ptrot} \mathU^{\step}_{\Ptrot-1} \cdots
    \Big( \frac{\partial}{\partial \theta^{\alpha}_p} \mathU^{\step}_p \Big) \cdots 
\mathU^{\step}_{1} \mathU_{0} \;,
\end{equation}
which in turn follows from the blocks derivative:
\begin{equation}
\frac{\partial}{\partial \theta^{\alpha}_p} \mathU^{\step}_p = 
    \left( \begin{array}{cc} \frac{\partial}{\partial \theta^{\alpha}_p} \U(\theta^x_p,\theta^z_p)  & \frac{\partial}{\partial \theta^{\alpha}_p} \V^*(\theta^x_p,\theta^z_p) \\ \frac{\partial}{\partial \theta^{\alpha}_p} \V(\theta^x_p,\theta^z_p) & \frac{\partial}{\partial \theta^{\alpha}_p} \U^*(\theta^x_p,\theta^z_p)  \end{array} \right) \;.
\end{equation}
More precisely, one can perform the calculation explicitly.
For the derivative with respect to $\theta^{x}_p$ the result is simple:
\begin{equation} \nonumber
\frac{\partial}{\partial \theta^{x}_p} \U(\theta^x_p,\theta^z_p) = 
    2i \U(\theta^x_p,\theta^z_p)  
 \;, \hspace{10mm} 
     \frac{\partial}{\partial \theta^{x}_p} \V(\theta^x_p,\theta^z_p) = 
    -2i \V(\theta^x_p,\theta^z_p)  
 \;.
\end{equation}
For the derivative with respect to $\theta^{z}_p$ we have:
\begin{align} 
    \frac{\partial}{\partial \theta^{z}_p} \U(\theta^x_p,\theta^z_p) 
    &= \nep^{+2i\theta^x_p }  \Big(
    \U_z \frac{\partial}{\partial \theta^{z}_p} \bfD_p 
    \U_z^{\dagger}
     + \V_z^* \frac{\partial}{\partial \theta^{z}_p} \bfD^*_p
     \V_z^{\transpose} \Big) \nonumber \vspace{4mm} \\
     \frac{\partial}{\partial \theta^{z}_p} \V(\theta^x_p,\theta^z_p) 
     &= \nep^{-2i\theta^x_p}  \Big(
    \V_z \frac{\partial}{\partial \theta^{z}_p} \bfD_p 
    \U_z^{\dagger}
     + \U_z^* \frac{\partial}{\partial \theta^{z}_p} \bfD^*_p 
     \V_z^{\transpose} \Big)
     \;, \nonumber
\end{align}
where the derivative 
$\frac{\partial}{\partial \theta^{z}_p} \bfD_p(\theta^z_p)$
follows directly from Eq.~\eqref{eqn:D_p}. 

\bibliography{biblio}

\begin{thebibliography}{47}%
\makeatletter
\providecommand \@ifxundefined [1]{%
 \@ifx{#1\undefined}
}%
\providecommand \@ifnum [1]{%
 \ifnum #1\expandafter \@firstoftwo
 \else \expandafter \@secondoftwo
 \fi
}%
\providecommand \@ifx [1]{%
 \ifx #1\expandafter \@firstoftwo
 \else \expandafter \@secondoftwo
 \fi
}%
\providecommand \natexlab [1]{#1}%
\providecommand \enquote  [1]{``#1''}%
\providecommand \bibnamefont  [1]{#1}%
\providecommand \bibfnamefont [1]{#1}%
\providecommand \citenamefont [1]{#1}%
\providecommand \href@noop [0]{\@secondoftwo}%
\providecommand \href [0]{\begingroup \@sanitize@url \@href}%
\providecommand \@href[1]{\@@startlink{#1}\@@href}%
\providecommand \@@href[1]{\endgroup#1\@@endlink}%
\providecommand \@sanitize@url [0]{\catcode `\\12\catcode `\$12\catcode
  `\&12\catcode `\#12\catcode `\^12\catcode `\_12\catcode `\%12\relax}%
\providecommand \@@startlink[1]{}%
\providecommand \@@endlink[0]{}%
\providecommand \url  [0]{\begingroup\@sanitize@url \@url }%
\providecommand \@url [1]{\endgroup\@href {#1}{\urlprefix }}%
\providecommand \urlprefix  [0]{URL }%
\providecommand \Eprint [0]{\href }%
\providecommand \doibase [0]{https://doi.org/}%
\providecommand \selectlanguage [0]{\@gobble}%
\providecommand \bibinfo  [0]{\@secondoftwo}%
\providecommand \bibfield  [0]{\@secondoftwo}%
\providecommand \translation [1]{[#1]}%
\providecommand \BibitemOpen [0]{}%
\providecommand \bibitemStop [0]{}%
\providecommand \bibitemNoStop [0]{.\EOS\space}%
\providecommand \EOS [0]{\spacefactor3000\relax}%
\providecommand \BibitemShut  [1]{\csname bibitem#1\endcsname}%
\let\auto@bib@innerbib\@empty
\bibitem [{\citenamefont {Abbas}\ \emph {et~al.}(2024)\citenamefont {Abbas},
  \citenamefont {Ambainis}, \citenamefont {Augustino}, \citenamefont
  {B{\"a}rtschi}, \citenamefont {Buhrman}, \citenamefont {Coffrin},
  \citenamefont {Cortiana}, \citenamefont {Dunjko}, \citenamefont {Egger},
  \citenamefont {Elmegreen} \emph {et~al.}}]{abbas2024challenges}%
  \BibitemOpen
  \bibfield  {author} {\bibinfo {author} {\bibfnamefont {A.}~\bibnamefont
  {Abbas}}, \bibinfo {author} {\bibfnamefont {A.}~\bibnamefont {Ambainis}},
  \bibinfo {author} {\bibfnamefont {B.}~\bibnamefont {Augustino}}, \bibinfo
  {author} {\bibfnamefont {A.}~\bibnamefont {B{\"a}rtschi}}, \bibinfo {author}
  {\bibfnamefont {H.}~\bibnamefont {Buhrman}}, \bibinfo {author} {\bibfnamefont
  {C.}~\bibnamefont {Coffrin}}, \bibinfo {author} {\bibfnamefont
  {G.}~\bibnamefont {Cortiana}}, \bibinfo {author} {\bibfnamefont
  {V.}~\bibnamefont {Dunjko}}, \bibinfo {author} {\bibfnamefont {D.~J.}\
  \bibnamefont {Egger}}, \bibinfo {author} {\bibfnamefont {B.~G.}\ \bibnamefont
  {Elmegreen}}, \emph {et~al.},\ }\href@noop {} {\bibfield  {journal} {\bibinfo
   {journal} {Nature Reviews Physics}\ ,\ \bibinfo {pages} {1}} (\bibinfo
  {year} {2024})}\BibitemShut {NoStop}%
\bibitem [{\citenamefont {Blekos}\ \emph {et~al.}(2024)\citenamefont {Blekos},
  \citenamefont {Brand}, \citenamefont {Ceschini}, \citenamefont {Chou},
  \citenamefont {Li}, \citenamefont {Pandya},\ and\ \citenamefont
  {Summer}}]{blekos2024review}%
  \BibitemOpen
  \bibfield  {author} {\bibinfo {author} {\bibfnamefont {K.}~\bibnamefont
  {Blekos}}, \bibinfo {author} {\bibfnamefont {D.}~\bibnamefont {Brand}},
  \bibinfo {author} {\bibfnamefont {A.}~\bibnamefont {Ceschini}}, \bibinfo
  {author} {\bibfnamefont {C.-H.}\ \bibnamefont {Chou}}, \bibinfo {author}
  {\bibfnamefont {R.-H.}\ \bibnamefont {Li}}, \bibinfo {author} {\bibfnamefont
  {K.}~\bibnamefont {Pandya}},\ and\ \bibinfo {author} {\bibfnamefont
  {A.}~\bibnamefont {Summer}},\ }\href@noop {} {\bibfield  {journal} {\bibinfo
  {journal} {Physics Reports}\ }\textbf {\bibinfo {volume} {1068}},\ \bibinfo
  {pages} {1} (\bibinfo {year} {2024})}\BibitemShut {NoStop}%
\bibitem [{\citenamefont {Klug}(2023)}]{klug2023quantum}%
  \BibitemOpen
  \bibfield  {author} {\bibinfo {author} {\bibfnamefont {F.}~\bibnamefont
  {Klug}},\ }\href@noop {} {\bibfield  {journal} {\bibinfo  {journal} {arXiv
  preprint arXiv:2312.13636}\ } (\bibinfo {year} {2023})}\BibitemShut {NoStop}%
\bibitem [{\citenamefont {Finnila}\ \emph {et~al.}(1994)\citenamefont
  {Finnila}, \citenamefont {Gomez}, \citenamefont {Sebenik}, \citenamefont
  {Stenson},\ and\ \citenamefont {Doll}}]{finnila_quantum_1994}%
  \BibitemOpen
  \bibfield  {author} {\bibinfo {author} {\bibfnamefont {A.~B.}\ \bibnamefont
  {Finnila}}, \bibinfo {author} {\bibfnamefont {M.~A.}\ \bibnamefont {Gomez}},
  \bibinfo {author} {\bibfnamefont {C.}~\bibnamefont {Sebenik}}, \bibinfo
  {author} {\bibfnamefont {C.}~\bibnamefont {Stenson}},\ and\ \bibinfo {author}
  {\bibfnamefont {J.~D.}\ \bibnamefont {Doll}},\ }\href
  {https://doi.org/https://doi.org/10.1016/0009-2614(94)00117-0} {\bibfield
  {journal} {\bibinfo  {journal} {Chemical Physics Letters}\ }\textbf {\bibinfo
  {volume} {219}},\ \bibinfo {pages} {343} (\bibinfo {year}
  {1994})}\BibitemShut {NoStop}%
\bibitem [{\citenamefont {Kadowaki}\ and\ \citenamefont
  {Nishimori}(1998)}]{kadowaki1998quantum}%
  \BibitemOpen
  \bibfield  {author} {\bibinfo {author} {\bibfnamefont {T.}~\bibnamefont
  {Kadowaki}}\ and\ \bibinfo {author} {\bibfnamefont {H.}~\bibnamefont
  {Nishimori}},\ }\href@noop {} {\bibfield  {journal} {\bibinfo  {journal}
  {Physical Review E}\ }\textbf {\bibinfo {volume} {58}},\ \bibinfo {pages}
  {5355} (\bibinfo {year} {1998})}\BibitemShut {NoStop}%
\bibitem [{\citenamefont {Santoro}\ \emph {et~al.}(2002)\citenamefont
  {Santoro}, \citenamefont {Martonak}, \citenamefont {Tosatti},\ and\
  \citenamefont {Car}}]{Santoro_SCI02}%
  \BibitemOpen
  \bibfield  {author} {\bibinfo {author} {\bibfnamefont {G.~E.}\ \bibnamefont
  {Santoro}}, \bibinfo {author} {\bibfnamefont {R.}~\bibnamefont {Martonak}},
  \bibinfo {author} {\bibfnamefont {E.}~\bibnamefont {Tosatti}},\ and\ \bibinfo
  {author} {\bibfnamefont {R.}~\bibnamefont {Car}},\ }\href
  {https://doi.org/10.1126/science.1068774} {\bibfield  {journal} {\bibinfo
  {journal} {Science}\ }\textbf {\bibinfo {volume} {295}},\ \bibinfo {pages}
  {2427} (\bibinfo {year} {2002})}\BibitemShut {NoStop}%
\bibitem [{\citenamefont {Santoro}\ and\ \citenamefont
  {Tosatti}(2006)}]{Santoro_2006}%
  \BibitemOpen
  \bibfield  {author} {\bibinfo {author} {\bibfnamefont {G.~E.}\ \bibnamefont
  {Santoro}}\ and\ \bibinfo {author} {\bibfnamefont {E.}~\bibnamefont
  {Tosatti}},\ }\href {https://doi.org/10.1088/0305-4470/39/36/R01} {\bibfield
  {journal} {\bibinfo  {journal} {Journal of Physics A: Mathematical and
  General}\ }\textbf {\bibinfo {volume} {39}},\ \bibinfo {pages} {R393}
  (\bibinfo {year} {2006})}\BibitemShut {NoStop}%
\bibitem [{\citenamefont {Albash}\ and\ \citenamefont
  {Lidar}(2018)}]{Albash_RMP18}%
  \BibitemOpen
  \bibfield  {author} {\bibinfo {author} {\bibfnamefont {T.}~\bibnamefont
  {Albash}}\ and\ \bibinfo {author} {\bibfnamefont {D.~A.}\ \bibnamefont
  {Lidar}},\ }\href@noop {} {\bibfield  {journal} {\bibinfo  {journal} {Rev.
  Mod. Phys.}\ }\textbf {\bibinfo {volume} {90}},\ \bibinfo {pages} {015002}
  (\bibinfo {year} {2018})}\BibitemShut {NoStop}%
\bibitem [{\citenamefont {Caneva}\ \emph {et~al.}(2007)\citenamefont {Caneva},
  \citenamefont {Fazio},\ and\ \citenamefont {Santoro}}]{Caneva_PRB2007}%
  \BibitemOpen
  \bibfield  {author} {\bibinfo {author} {\bibfnamefont {T.}~\bibnamefont
  {Caneva}}, \bibinfo {author} {\bibfnamefont {R.}~\bibnamefont {Fazio}},\ and\
  \bibinfo {author} {\bibfnamefont {G.~E.}\ \bibnamefont {Santoro}},\ }\href
  {https://doi.org/10.1103/PhysRevB.76.144427} {\bibfield  {journal} {\bibinfo
  {journal} {Phys. Rev. B}\ }\textbf {\bibinfo {volume} {76}},\ \bibinfo
  {pages} {144427} (\bibinfo {year} {2007})}\BibitemShut {NoStop}%
\bibitem [{\citenamefont {Knysh}(2016)}]{Knysh_NatComm16}%
  \BibitemOpen
  \bibfield  {author} {\bibinfo {author} {\bibfnamefont {S.}~\bibnamefont
  {Knysh}},\ }\href@noop {} {\bibfield  {journal} {\bibinfo  {journal} {Nature
  Communications}\ }\textbf {\bibinfo {volume} {7}},\ \bibinfo {pages} {12370}
  (\bibinfo {year} {2016})}\BibitemShut {NoStop}%
\bibitem [{\citenamefont {Roberts}\ \emph {et~al.}(2020)\citenamefont
  {Roberts}, \citenamefont {Cincio}, \citenamefont {Saxena}, \citenamefont
  {Petukhov},\ and\ \citenamefont {Knysh}}]{Knysh_PRA2020}%
  \BibitemOpen
  \bibfield  {author} {\bibinfo {author} {\bibfnamefont {D.}~\bibnamefont
  {Roberts}}, \bibinfo {author} {\bibfnamefont {L.}~\bibnamefont {Cincio}},
  \bibinfo {author} {\bibfnamefont {A.}~\bibnamefont {Saxena}}, \bibinfo
  {author} {\bibfnamefont {A.}~\bibnamefont {Petukhov}},\ and\ \bibinfo
  {author} {\bibfnamefont {S.}~\bibnamefont {Knysh}},\ }\href
  {https://doi.org/10.1103/PhysRevA.101.042317} {\bibfield  {journal} {\bibinfo
   {journal} {Phys. Rev. A}\ }\textbf {\bibinfo {volume} {101}},\ \bibinfo
  {pages} {042317} (\bibinfo {year} {2020})}\BibitemShut {NoStop}%
\bibitem [{\citenamefont {Bapst}\ \emph {et~al.}(2013)\citenamefont {Bapst},
  \citenamefont {Foini}, \citenamefont {Krzakala}, \citenamefont {Semerjian},\
  and\ \citenamefont {Zamponi}}]{Zamponi_QA:review}%
  \BibitemOpen
  \bibfield  {author} {\bibinfo {author} {\bibfnamefont {V.}~\bibnamefont
  {Bapst}}, \bibinfo {author} {\bibfnamefont {L.}~\bibnamefont {Foini}},
  \bibinfo {author} {\bibfnamefont {F.}~\bibnamefont {Krzakala}}, \bibinfo
  {author} {\bibfnamefont {G.}~\bibnamefont {Semerjian}},\ and\ \bibinfo
  {author} {\bibfnamefont {F.}~\bibnamefont {Zamponi}},\ }\href@noop {}
  {\bibfield  {journal} {\bibinfo  {journal} {Phys. Rep.}\ }\textbf {\bibinfo
  {volume} {523}},\ \bibinfo {pages} {127} (\bibinfo {year}
  {2013})}\BibitemShut {NoStop}%
\bibitem [{\citenamefont {Matsuura}\ \emph {et~al.}(2021)\citenamefont
  {Matsuura}, \citenamefont {Buck}, \citenamefont {Senicourt},\ and\
  \citenamefont {Zaribafiyan}}]{Matsuura_PRA2021}%
  \BibitemOpen
  \bibfield  {author} {\bibinfo {author} {\bibfnamefont {S.}~\bibnamefont
  {Matsuura}}, \bibinfo {author} {\bibfnamefont {S.}~\bibnamefont {Buck}},
  \bibinfo {author} {\bibfnamefont {V.}~\bibnamefont {Senicourt}},\ and\
  \bibinfo {author} {\bibfnamefont {A.}~\bibnamefont {Zaribafiyan}},\ }\href
  {https://doi.org/10.1103/PhysRevA.103.052435} {\bibfield  {journal} {\bibinfo
   {journal} {Phys. Rev. A}\ }\textbf {\bibinfo {volume} {103}},\ \bibinfo
  {pages} {052435} (\bibinfo {year} {2021})}\BibitemShut {NoStop}%
\bibitem [{\citenamefont {Côté}\ \emph {et~al.}(2023)\citenamefont {Côté},
  \citenamefont {Sauvage}, \citenamefont {Larocca}, \citenamefont {Jonsson},
  \citenamefont {Cincio},\ and\ \citenamefont {Albash}}]{Cote_2023}%
  \BibitemOpen
  \bibfield  {author} {\bibinfo {author} {\bibfnamefont {J.}~\bibnamefont
  {Côté}}, \bibinfo {author} {\bibfnamefont {F.}~\bibnamefont {Sauvage}},
  \bibinfo {author} {\bibfnamefont {M.}~\bibnamefont {Larocca}}, \bibinfo
  {author} {\bibfnamefont {M.}~\bibnamefont {Jonsson}}, \bibinfo {author}
  {\bibfnamefont {L.}~\bibnamefont {Cincio}},\ and\ \bibinfo {author}
  {\bibfnamefont {T.}~\bibnamefont {Albash}},\ }\href
  {https://doi.org/10.1088/2058-9565/acfbaa} {\bibfield  {journal} {\bibinfo
  {journal} {Quantum Science and Technology}\ }\textbf {\bibinfo {volume}
  {8}},\ \bibinfo {pages} {045033} (\bibinfo {year} {2023})}\BibitemShut
  {NoStop}%
\bibitem [{\citenamefont {Quiroz}(2019)}]{Quiroz_PRA2019}%
  \BibitemOpen
  \bibfield  {author} {\bibinfo {author} {\bibfnamefont {G.}~\bibnamefont
  {Quiroz}},\ }\href {https://doi.org/10.1103/PhysRevA.99.062306} {\bibfield
  {journal} {\bibinfo  {journal} {Phys. Rev. A}\ }\textbf {\bibinfo {volume}
  {99}},\ \bibinfo {pages} {062306} (\bibinfo {year} {2019})}\BibitemShut
  {NoStop}%
\bibitem [{\citenamefont {Hegde}\ \emph {et~al.}(2022)\citenamefont {Hegde},
  \citenamefont {Passarelli}, \citenamefont {Scocco},\ and\ \citenamefont
  {Lucignano}}]{Lucignano_PRA2022}%
  \BibitemOpen
  \bibfield  {author} {\bibinfo {author} {\bibfnamefont {P.~R.}\ \bibnamefont
  {Hegde}}, \bibinfo {author} {\bibfnamefont {G.}~\bibnamefont {Passarelli}},
  \bibinfo {author} {\bibfnamefont {A.}~\bibnamefont {Scocco}},\ and\ \bibinfo
  {author} {\bibfnamefont {P.}~\bibnamefont {Lucignano}},\ }\href
  {https://doi.org/10.1103/PhysRevA.105.012612} {\bibfield  {journal} {\bibinfo
   {journal} {Phys. Rev. A}\ }\textbf {\bibinfo {volume} {105}},\ \bibinfo
  {pages} {012612} (\bibinfo {year} {2022})}\BibitemShut {NoStop}%
\bibitem [{\citenamefont {Passarelli}\ \emph {et~al.}(2019)\citenamefont
  {Passarelli}, \citenamefont {Cataudella},\ and\ \citenamefont
  {Lucignano}}]{Passarelli_PRB2019}%
  \BibitemOpen
  \bibfield  {author} {\bibinfo {author} {\bibfnamefont {G.}~\bibnamefont
  {Passarelli}}, \bibinfo {author} {\bibfnamefont {V.}~\bibnamefont
  {Cataudella}},\ and\ \bibinfo {author} {\bibfnamefont {P.}~\bibnamefont
  {Lucignano}},\ }\href {https://doi.org/10.1103/PhysRevB.100.024302}
  {\bibfield  {journal} {\bibinfo  {journal} {Phys. Rev. B}\ }\textbf {\bibinfo
  {volume} {100}},\ \bibinfo {pages} {024302} (\bibinfo {year}
  {2019})}\BibitemShut {NoStop}%
\bibitem [{\citenamefont {Caneva}\ \emph {et~al.}(2011)\citenamefont {Caneva},
  \citenamefont {Calarco},\ and\ \citenamefont {Montangero}}]{Caneva_PRA2011}%
  \BibitemOpen
  \bibfield  {author} {\bibinfo {author} {\bibfnamefont {T.}~\bibnamefont
  {Caneva}}, \bibinfo {author} {\bibfnamefont {T.}~\bibnamefont {Calarco}},\
  and\ \bibinfo {author} {\bibfnamefont {S.}~\bibnamefont {Montangero}},\
  }\href {https://doi.org/10.1103/PhysRevA.84.022326} {\bibfield  {journal}
  {\bibinfo  {journal} {Phys. Rev. A}\ }\textbf {\bibinfo {volume} {84}},\
  \bibinfo {pages} {022326} (\bibinfo {year} {2011})}\BibitemShut {NoStop}%
\bibitem [{\citenamefont {Rach}\ \emph {et~al.}(2015)\citenamefont {Rach},
  \citenamefont {M\"uller}, \citenamefont {Calarco},\ and\ \citenamefont
  {Montangero}}]{Montangero_dCRAB_PRA2015}%
  \BibitemOpen
  \bibfield  {author} {\bibinfo {author} {\bibfnamefont {N.}~\bibnamefont
  {Rach}}, \bibinfo {author} {\bibfnamefont {M.~M.}\ \bibnamefont {M\"uller}},
  \bibinfo {author} {\bibfnamefont {T.}~\bibnamefont {Calarco}},\ and\ \bibinfo
  {author} {\bibfnamefont {S.}~\bibnamefont {Montangero}},\ }\href
  {https://doi.org/10.1103/PhysRevA.92.062343} {\bibfield  {journal} {\bibinfo
  {journal} {Phys. Rev. A}\ }\textbf {\bibinfo {volume} {92}},\ \bibinfo
  {pages} {062343} (\bibinfo {year} {2015})}\BibitemShut {NoStop}%
\bibitem [{\citenamefont {Barends}\ \emph {et~al.}(2016)\citenamefont
  {Barends}, \citenamefont {Shabani}, \citenamefont {Lamata}, \citenamefont
  {Kelly}, \citenamefont {Mezzacapo}, \citenamefont {Heras}, \citenamefont
  {Babbush}, \citenamefont {Fowler}, \citenamefont {Campbell}, \citenamefont
  {Chen}, \citenamefont {Chen}, \citenamefont {Chiaro}, \citenamefont
  {Dunsworth}, \citenamefont {Jeffrey}, \citenamefont {Lucero}, \citenamefont
  {Megrant}, \citenamefont {Mutus}, \citenamefont {Neeley}, \citenamefont
  {Neill}, \citenamefont {O'Malley}, \citenamefont {Quintana}, \citenamefont
  {Roushan}, \citenamefont {Sank}, \citenamefont {Vainsencher}, \citenamefont
  {Wenner}, \citenamefont {White}, \citenamefont {Solano}, \citenamefont
  {Neven},\ and\ \citenamefont {Martinis}}]{Nature_dQA}%
  \BibitemOpen
  \bibfield  {author} {\bibinfo {author} {\bibfnamefont {R.}~\bibnamefont
  {Barends}}, \bibinfo {author} {\bibfnamefont {A.}~\bibnamefont {Shabani}},
  \bibinfo {author} {\bibfnamefont {L.}~\bibnamefont {Lamata}}, \bibinfo
  {author} {\bibfnamefont {J.}~\bibnamefont {Kelly}}, \bibinfo {author}
  {\bibfnamefont {A.}~\bibnamefont {Mezzacapo}}, \bibinfo {author}
  {\bibfnamefont {U.~L.}\ \bibnamefont {Heras}}, \bibinfo {author}
  {\bibfnamefont {R.}~\bibnamefont {Babbush}}, \bibinfo {author} {\bibfnamefont
  {A.~G.}\ \bibnamefont {Fowler}}, \bibinfo {author} {\bibfnamefont
  {B.}~\bibnamefont {Campbell}}, \bibinfo {author} {\bibfnamefont
  {Y.}~\bibnamefont {Chen}}, \bibinfo {author} {\bibfnamefont {Z.}~\bibnamefont
  {Chen}}, \bibinfo {author} {\bibfnamefont {B.}~\bibnamefont {Chiaro}},
  \bibinfo {author} {\bibfnamefont {A.}~\bibnamefont {Dunsworth}}, \bibinfo
  {author} {\bibfnamefont {E.}~\bibnamefont {Jeffrey}}, \bibinfo {author}
  {\bibfnamefont {E.}~\bibnamefont {Lucero}}, \bibinfo {author} {\bibfnamefont
  {A.}~\bibnamefont {Megrant}}, \bibinfo {author} {\bibfnamefont {J.~Y.}\
  \bibnamefont {Mutus}}, \bibinfo {author} {\bibfnamefont {M.}~\bibnamefont
  {Neeley}}, \bibinfo {author} {\bibfnamefont {C.}~\bibnamefont {Neill}},
  \bibinfo {author} {\bibfnamefont {P.~J.~J.}\ \bibnamefont {O'Malley}},
  \bibinfo {author} {\bibfnamefont {C.}~\bibnamefont {Quintana}}, \bibinfo
  {author} {\bibfnamefont {P.}~\bibnamefont {Roushan}}, \bibinfo {author}
  {\bibfnamefont {D.}~\bibnamefont {Sank}}, \bibinfo {author} {\bibfnamefont
  {A.}~\bibnamefont {Vainsencher}}, \bibinfo {author} {\bibfnamefont
  {J.}~\bibnamefont {Wenner}}, \bibinfo {author} {\bibfnamefont {T.~C.}\
  \bibnamefont {White}}, \bibinfo {author} {\bibfnamefont {E.}~\bibnamefont
  {Solano}}, \bibinfo {author} {\bibfnamefont {H.}~\bibnamefont {Neven}},\ and\
  \bibinfo {author} {\bibfnamefont {J.~M.}\ \bibnamefont {Martinis}},\
  }\href@noop {} {\bibfield  {journal} {\bibinfo  {journal} {Nature}\ }\textbf
  {\bibinfo {volume} {534}},\ \bibinfo {pages} {222} (\bibinfo {year}
  {2016})}\BibitemShut {NoStop}%
\bibitem [{\citenamefont {Mbeng}\ \emph
  {et~al.}(2019{\natexlab{a}})\citenamefont {Mbeng}, \citenamefont {Arceci},\
  and\ \citenamefont {Santoro}}]{Mbeng_dQA_PRB2019}%
  \BibitemOpen
  \bibfield  {author} {\bibinfo {author} {\bibfnamefont {G.~B.}\ \bibnamefont
  {Mbeng}}, \bibinfo {author} {\bibfnamefont {L.}~\bibnamefont {Arceci}},\ and\
  \bibinfo {author} {\bibfnamefont {G.~E.}\ \bibnamefont {Santoro}},\ }\href
  {https://doi.org/10.1103/PhysRevB.100.224201} {\bibfield  {journal} {\bibinfo
   {journal} {Phys. Rev. B}\ }\textbf {\bibinfo {volume} {100}},\ \bibinfo
  {pages} {224201} (\bibinfo {year} {2019}{\natexlab{a}})}\BibitemShut
  {NoStop}%
\bibitem [{\citenamefont {Farhi}\ \emph {et~al.}(2014)\citenamefont {Farhi},
  \citenamefont {Goldstone},\ and\ \citenamefont
  {Gutmann}}]{farhi_quantum_2014}%
  \BibitemOpen
  \bibfield  {author} {\bibinfo {author} {\bibfnamefont {E.}~\bibnamefont
  {Farhi}}, \bibinfo {author} {\bibfnamefont {J.}~\bibnamefont {Goldstone}},\
  and\ \bibinfo {author} {\bibfnamefont {S.}~\bibnamefont {Gutmann}},\ }\href
  {https://doi.org/10.48550/arXiv.1411.4028} {\bibinfo {title} {A {Quantum}
  {Approximate} {Optimization} {Algorithm}}} (\bibinfo {year} {2014}),\
  \bibinfo {note} {arXiv:1411.4028 [quant-ph]}\BibitemShut {NoStop}%
\bibitem [{\citenamefont {Cerezo}\ \emph {et~al.}(2021)\citenamefont {Cerezo},
  \citenamefont {Arrasmith}, \citenamefont {Babbush}, \citenamefont {Benjamin},
  \citenamefont {Endo}, \citenamefont {Fujii}, \citenamefont {McClean},
  \citenamefont {Mitarai}, \citenamefont {Yuan}, \citenamefont {Cincio},\ and\
  \citenamefont {Coles}}]{cerezo_variational_2021}%
  \BibitemOpen
  \bibfield  {author} {\bibinfo {author} {\bibfnamefont {M.}~\bibnamefont
  {Cerezo}}, \bibinfo {author} {\bibfnamefont {A.}~\bibnamefont {Arrasmith}},
  \bibinfo {author} {\bibfnamefont {R.}~\bibnamefont {Babbush}}, \bibinfo
  {author} {\bibfnamefont {S.~C.}\ \bibnamefont {Benjamin}}, \bibinfo {author}
  {\bibfnamefont {S.}~\bibnamefont {Endo}}, \bibinfo {author} {\bibfnamefont
  {K.}~\bibnamefont {Fujii}}, \bibinfo {author} {\bibfnamefont {J.~R.}\
  \bibnamefont {McClean}}, \bibinfo {author} {\bibfnamefont {K.}~\bibnamefont
  {Mitarai}}, \bibinfo {author} {\bibfnamefont {X.}~\bibnamefont {Yuan}},
  \bibinfo {author} {\bibfnamefont {L.}~\bibnamefont {Cincio}},\ and\ \bibinfo
  {author} {\bibfnamefont {P.~J.}\ \bibnamefont {Coles}},\ }\href
  {https://doi.org/10.1038/s42254-021-00348-9} {\bibfield  {journal} {\bibinfo
  {journal} {Nature Reviews Physics}\ }\textbf {\bibinfo {volume} {3}},\
  \bibinfo {pages} {625} (\bibinfo {year} {2021})}\BibitemShut {NoStop}%
\bibitem [{\citenamefont {Pecci}\ \emph {et~al.}(2024)\citenamefont {Pecci},
  \citenamefont {Wang}, \citenamefont {Torta}, \citenamefont {Mbeng},\ and\
  \citenamefont {Santoro}}]{pecci2024beyond}%
  \BibitemOpen
  \bibfield  {author} {\bibinfo {author} {\bibfnamefont {G.}~\bibnamefont
  {Pecci}}, \bibinfo {author} {\bibfnamefont {R.}~\bibnamefont {Wang}},
  \bibinfo {author} {\bibfnamefont {P.}~\bibnamefont {Torta}}, \bibinfo
  {author} {\bibfnamefont {G.~B.}\ \bibnamefont {Mbeng}},\ and\ \bibinfo
  {author} {\bibfnamefont {G.}~\bibnamefont {Santoro}},\ }\href
  {https://doi.org/10.1088/2058-9565/ad60f2} {\bibfield  {journal} {\bibinfo
  {journal} {Quantum Science and Technology}\ }\textbf {\bibinfo {volume}
  {9}},\ \bibinfo {pages} {045013} (\bibinfo {year} {2024})}\BibitemShut
  {NoStop}%
\bibitem [{\citenamefont {D'Alessandro}(2007)}]{Dalessandro2007}%
  \BibitemOpen
  \bibfield  {author} {\bibinfo {author} {\bibfnamefont {D.}~\bibnamefont
  {D'Alessandro}},\ }\href@noop {} {\emph {\bibinfo {title} {Introduction to
  quantum control and dynamics}}}\ (\bibinfo  {publisher} {Chapman and
  Hall/CRC},\ \bibinfo {year} {2007})\BibitemShut {NoStop}%
\bibitem [{\citenamefont {Margolus}\ and\ \citenamefont
  {Levitin}(1998)}]{MARGOLUS1998188}%
  \BibitemOpen
  \bibfield  {author} {\bibinfo {author} {\bibfnamefont {N.}~\bibnamefont
  {Margolus}}\ and\ \bibinfo {author} {\bibfnamefont {L.~B.}\ \bibnamefont
  {Levitin}},\ }\href
  {https://doi.org/https://doi.org/10.1016/S0167-2789(98)00054-2} {\bibfield
  {journal} {\bibinfo  {journal} {Physica D: Nonlinear Phenomena}\ }\textbf
  {\bibinfo {volume} {120}},\ \bibinfo {pages} {188} (\bibinfo {year}
  {1998})},\ \bibinfo {note} {proceedings of the Fourth Workshop on Physics and
  Consumption}\BibitemShut {NoStop}%
\bibitem [{\citenamefont {Larocca}\ \emph {et~al.}(2022)\citenamefont
  {Larocca}, \citenamefont {Czarnik}, \citenamefont {Sharma}, \citenamefont
  {Muraleedharan}, \citenamefont {Coles},\ and\ \citenamefont
  {Cerezo}}]{Larocca2022diagnosingbarren}%
  \BibitemOpen
  \bibfield  {author} {\bibinfo {author} {\bibfnamefont {M.}~\bibnamefont
  {Larocca}}, \bibinfo {author} {\bibfnamefont {P.}~\bibnamefont {Czarnik}},
  \bibinfo {author} {\bibfnamefont {K.}~\bibnamefont {Sharma}}, \bibinfo
  {author} {\bibfnamefont {G.}~\bibnamefont {Muraleedharan}}, \bibinfo {author}
  {\bibfnamefont {P.~J.}\ \bibnamefont {Coles}},\ and\ \bibinfo {author}
  {\bibfnamefont {M.}~\bibnamefont {Cerezo}},\ }\href
  {https://doi.org/10.22331/q-2022-09-29-824} {\bibfield  {journal} {\bibinfo
  {journal} {{Quantum}}\ }\textbf {\bibinfo {volume} {6}},\ \bibinfo {pages}
  {824} (\bibinfo {year} {2022})}\BibitemShut {NoStop}%
\bibitem [{\citenamefont {Mbeng}\ \emph
  {et~al.}(2019{\natexlab{b}})\citenamefont {Mbeng}, \citenamefont {Fazio},\
  and\ \citenamefont {Santoro}}]{mbeng2019optimal}%
  \BibitemOpen
  \bibfield  {author} {\bibinfo {author} {\bibfnamefont {G.~B.}\ \bibnamefont
  {Mbeng}}, \bibinfo {author} {\bibfnamefont {R.}~\bibnamefont {Fazio}},\ and\
  \bibinfo {author} {\bibfnamefont {G.~E.}\ \bibnamefont {Santoro}},\
  }\href@noop {} {\bibinfo {title} {Optimal quantum control with digitized
  quantum annealing}} (\bibinfo {year} {2019}{\natexlab{b}}),\ \Eprint
  {https://arxiv.org/abs/1911.12259} {arXiv:1911.12259 [quant-ph]} \BibitemShut
  {NoStop}%
\bibitem [{\citenamefont {Niu}\ \emph {et~al.}(2019)\citenamefont {Niu},
  \citenamefont {Lu},\ and\ \citenamefont
  {Chuang}}]{niu2019optimizingqaoasuccessprobability}%
  \BibitemOpen
  \bibfield  {author} {\bibinfo {author} {\bibfnamefont {M.~Y.}\ \bibnamefont
  {Niu}}, \bibinfo {author} {\bibfnamefont {S.}~\bibnamefont {Lu}},\ and\
  \bibinfo {author} {\bibfnamefont {I.~L.}\ \bibnamefont {Chuang}},\ }\href
  {https://arxiv.org/abs/1905.12134} {\bibinfo {title} {Optimizing qaoa:
  Success probability and runtime dependence on circuit depth}} (\bibinfo
  {year} {2019}),\ \Eprint {https://arxiv.org/abs/1905.12134} {arXiv:1905.12134
  [quant-ph]} \BibitemShut {NoStop}%
\bibitem [{\citenamefont {Doria}\ \emph {et~al.}(2011)\citenamefont {Doria},
  \citenamefont {Calarco},\ and\ \citenamefont {Montangero}}]{Doria_PRL2011}%
  \BibitemOpen
  \bibfield  {author} {\bibinfo {author} {\bibfnamefont {P.}~\bibnamefont
  {Doria}}, \bibinfo {author} {\bibfnamefont {T.}~\bibnamefont {Calarco}},\
  and\ \bibinfo {author} {\bibfnamefont {S.}~\bibnamefont {Montangero}},\
  }\href {https://doi.org/10.1103/PhysRevLett.106.190501} {\bibfield  {journal}
  {\bibinfo  {journal} {Phys. Rev. Lett.}\ }\textbf {\bibinfo {volume} {106}},\
  \bibinfo {pages} {190501} (\bibinfo {year} {2011})}\BibitemShut {NoStop}%
\bibitem [{\citenamefont {Koch}\ \emph {et~al.}(2022)\citenamefont {Koch},
  \citenamefont {Boscain}, \citenamefont {Calarco}, \citenamefont {Dirr},
  \citenamefont {Filipp}, \citenamefont {Glaser}, \citenamefont {Kosloff},
  \citenamefont {Montangero}, \citenamefont {Schulte-Herbr{\"u}ggen},
  \citenamefont {Sugny},\ and\ \citenamefont {Wilhelm}}]{koch_quantum_2022}%
  \BibitemOpen
  \bibfield  {author} {\bibinfo {author} {\bibfnamefont {C.~P.}\ \bibnamefont
  {Koch}}, \bibinfo {author} {\bibfnamefont {U.}~\bibnamefont {Boscain}},
  \bibinfo {author} {\bibfnamefont {T.}~\bibnamefont {Calarco}}, \bibinfo
  {author} {\bibfnamefont {G.}~\bibnamefont {Dirr}}, \bibinfo {author}
  {\bibfnamefont {S.}~\bibnamefont {Filipp}}, \bibinfo {author} {\bibfnamefont
  {S.~J.}\ \bibnamefont {Glaser}}, \bibinfo {author} {\bibfnamefont
  {R.}~\bibnamefont {Kosloff}}, \bibinfo {author} {\bibfnamefont
  {S.}~\bibnamefont {Montangero}}, \bibinfo {author} {\bibfnamefont
  {T.}~\bibnamefont {Schulte-Herbr{\"u}ggen}}, \bibinfo {author} {\bibfnamefont
  {D.}~\bibnamefont {Sugny}},\ and\ \bibinfo {author} {\bibfnamefont {F.~K.}\
  \bibnamefont {Wilhelm}},\ }\href@noop {} {\bibfield  {journal} {\bibinfo
  {journal} {EPJ Quantum Technology}\ }\textbf {\bibinfo {volume} {9}},\
  \bibinfo {pages} {19} (\bibinfo {year} {2022})}\BibitemShut {NoStop}%
\bibitem [{\citenamefont {Gu\'ery-Odelin}\ \emph {et~al.}(2019)\citenamefont
  {Gu\'ery-Odelin}, \citenamefont {Ruschhaupt}, \citenamefont {Kiely},
  \citenamefont {Torrontegui}, \citenamefont {Mart\'{\i}nez-Garaot},\ and\
  \citenamefont {Muga}}]{STA_Review_RMP2019}%
  \BibitemOpen
  \bibfield  {author} {\bibinfo {author} {\bibfnamefont {D.}~\bibnamefont
  {Gu\'ery-Odelin}}, \bibinfo {author} {\bibfnamefont {A.}~\bibnamefont
  {Ruschhaupt}}, \bibinfo {author} {\bibfnamefont {A.}~\bibnamefont {Kiely}},
  \bibinfo {author} {\bibfnamefont {E.}~\bibnamefont {Torrontegui}}, \bibinfo
  {author} {\bibfnamefont {S.}~\bibnamefont {Mart\'{\i}nez-Garaot}},\ and\
  \bibinfo {author} {\bibfnamefont {J.~G.}\ \bibnamefont {Muga}},\ }\href
  {https://doi.org/10.1103/RevModPhys.91.045001} {\bibfield  {journal}
  {\bibinfo  {journal} {Rev. Mod. Phys.}\ }\textbf {\bibinfo {volume} {91}},\
  \bibinfo {pages} {045001} (\bibinfo {year} {2019})}\BibitemShut {NoStop}%
\bibitem [{\citenamefont {Torrontegui}\ \emph {et~al.}(2013)\citenamefont
  {Torrontegui}, \citenamefont {Ibáñez}, \citenamefont {Martínez-Garaot},
  \citenamefont {Modugno}, \citenamefont {{del Campo}}, \citenamefont
  {Guéry-Odelin}, \citenamefont {Ruschhaupt}, \citenamefont {Chen},\ and\
  \citenamefont {Muga}}]{TORRONTEGUI2013117}%
  \BibitemOpen
  \bibfield  {author} {\bibinfo {author} {\bibfnamefont {E.}~\bibnamefont
  {Torrontegui}}, \bibinfo {author} {\bibfnamefont {S.}~\bibnamefont
  {Ibáñez}}, \bibinfo {author} {\bibfnamefont {S.}~\bibnamefont
  {Martínez-Garaot}}, \bibinfo {author} {\bibfnamefont {M.}~\bibnamefont
  {Modugno}}, \bibinfo {author} {\bibfnamefont {A.}~\bibnamefont {{del
  Campo}}}, \bibinfo {author} {\bibfnamefont {D.}~\bibnamefont
  {Guéry-Odelin}}, \bibinfo {author} {\bibfnamefont {A.}~\bibnamefont
  {Ruschhaupt}}, \bibinfo {author} {\bibfnamefont {X.}~\bibnamefont {Chen}},\
  and\ \bibinfo {author} {\bibfnamefont {J.~G.}\ \bibnamefont {Muga}},\ }in\
  \href {https://doi.org/https://doi.org/10.1016/B978-0-12-408090-4.00002-5}
  {\emph {\bibinfo {booktitle} {Advances in Atomic, Molecular, and Optical
  Physics}}},\ \bibinfo {series} {Advances In Atomic, Molecular, and Optical
  Physics}, Vol.~\bibinfo {volume} {62},\ \bibinfo {editor} {edited by\
  \bibinfo {editor} {\bibfnamefont {E.}~\bibnamefont {Arimondo}}, \bibinfo
  {editor} {\bibfnamefont {P.~R.}\ \bibnamefont {Berman}},\ and\ \bibinfo
  {editor} {\bibfnamefont {C.~C.}\ \bibnamefont {Lin}}}\ (\bibinfo  {publisher}
  {Academic Press},\ \bibinfo {year} {2013})\ pp.\ \bibinfo {pages}
  {117--169}\BibitemShut {NoStop}%
\bibitem [{\citenamefont {Mbeng}\ \emph {et~al.}(2024)\citenamefont {Mbeng},
  \citenamefont {Russomanno},\ and\ \citenamefont
  {Santoro}}]{mbeng2024quantum}%
  \BibitemOpen
  \bibfield  {author} {\bibinfo {author} {\bibfnamefont {G.~B.}\ \bibnamefont
  {Mbeng}}, \bibinfo {author} {\bibfnamefont {A.}~\bibnamefont {Russomanno}},\
  and\ \bibinfo {author} {\bibfnamefont {G.~E.}\ \bibnamefont {Santoro}},\
  }\href@noop {} {\bibfield  {journal} {\bibinfo  {journal} {SciPost Physics
  Lecture Notes}\ ,\ \bibinfo {pages} {082}} (\bibinfo {year}
  {2024})}\BibitemShut {NoStop}%
\bibitem [{\citenamefont {Nocedal}\ and\ \citenamefont
  {Wright}(1999)}]{nocedal1999numerical}%
  \BibitemOpen
  \bibfield  {author} {\bibinfo {author} {\bibfnamefont {J.}~\bibnamefont
  {Nocedal}}\ and\ \bibinfo {author} {\bibfnamefont {S.~J.}\ \bibnamefont
  {Wright}},\ }\href@noop {} {\emph {\bibinfo {title} {Numerical
  optimization}}}\ (\bibinfo  {publisher} {Springer},\ \bibinfo {year}
  {1999})\BibitemShut {NoStop}%
\bibitem [{\citenamefont {{Farhi}}\ \emph {et~al.}(2014)\citenamefont
  {{Farhi}}, \citenamefont {{Goldstone}},\ and\ \citenamefont
  {{Gutmann}}}]{Farhi_arXiv2014}%
  \BibitemOpen
  \bibfield  {author} {\bibinfo {author} {\bibfnamefont {E.}~\bibnamefont
  {{Farhi}}}, \bibinfo {author} {\bibfnamefont {J.}~\bibnamefont
  {{Goldstone}}},\ and\ \bibinfo {author} {\bibfnamefont {S.}~\bibnamefont
  {{Gutmann}}},\ }\href@noop {} {\bibinfo {title} {A quantum approximate
  optimization algorithm}} (\bibinfo {year} {2014}),\ \Eprint
  {https://arxiv.org/abs/1411.4028} {arXiv:1411.4028 [quant-ph]} \BibitemShut
  {NoStop}%
\bibitem [{\citenamefont {Arezzo}\ \emph {et~al.}(2025)\citenamefont {Arezzo},
  \citenamefont {Wang}, \citenamefont {Thengil}, \citenamefont {Pecci},\ and\
  \citenamefont {Santoro}}]{Arezzo_inpreparation}%
  \BibitemOpen
  \bibfield  {author} {\bibinfo {author} {\bibfnamefont {V.~R.}\ \bibnamefont
  {Arezzo}}, \bibinfo {author} {\bibfnamefont {R.}~\bibnamefont {Wang}},
  \bibinfo {author} {\bibfnamefont {K.}~\bibnamefont {Thengil}}, \bibinfo
  {author} {\bibfnamefont {G.}~\bibnamefont {Pecci}},\ and\ \bibinfo {author}
  {\bibfnamefont {G.~E.}\ \bibnamefont {Santoro}}} (\bibinfo {year} {2025}),\
  \bibinfo {note} {(in preparation)}\BibitemShut {NoStop}%
\bibitem [{\citenamefont {Albertini}\ and\ \citenamefont
  {D'Alessandro}(2002)}]{ALBERTINI2002213}%
  \BibitemOpen
  \bibfield  {author} {\bibinfo {author} {\bibfnamefont {F.}~\bibnamefont
  {Albertini}}\ and\ \bibinfo {author} {\bibfnamefont {D.}~\bibnamefont
  {D'Alessandro}},\ }\href
  {https://doi.org/https://doi.org/10.1016/S0024-3795(02)00290-2} {\bibfield
  {journal} {\bibinfo  {journal} {Linear Algebra and its Applications}\
  }\textbf {\bibinfo {volume} {350}},\ \bibinfo {pages} {213} (\bibinfo {year}
  {2002})}\BibitemShut {NoStop}%
\bibitem [{\citenamefont {Giovannetti}\ \emph {et~al.}(2003)\citenamefont
  {Giovannetti}, \citenamefont {Lloyd},\ and\ \citenamefont
  {Maccone}}]{giovannetti_qsl}%
  \BibitemOpen
  \bibfield  {author} {\bibinfo {author} {\bibfnamefont {V.}~\bibnamefont
  {Giovannetti}}, \bibinfo {author} {\bibfnamefont {S.}~\bibnamefont {Lloyd}},\
  and\ \bibinfo {author} {\bibfnamefont {L.}~\bibnamefont {Maccone}},\ }\href
  {https://doi.org/10.1103/PhysRevA.67.052109} {\bibfield  {journal} {\bibinfo
  {journal} {Phys. Rev. A}\ }\textbf {\bibinfo {volume} {67}},\ \bibinfo
  {pages} {052109} (\bibinfo {year} {2003})}\BibitemShut {NoStop}%
\bibitem [{\citenamefont {Levitin}\ and\ \citenamefont
  {Toffoli}(2009)}]{levitin_qsl}%
  \BibitemOpen
  \bibfield  {author} {\bibinfo {author} {\bibfnamefont {L.~B.}\ \bibnamefont
  {Levitin}}\ and\ \bibinfo {author} {\bibfnamefont {T.}~\bibnamefont
  {Toffoli}},\ }\href {https://doi.org/10.1103/PhysRevLett.103.160502}
  {\bibfield  {journal} {\bibinfo  {journal} {Phys. Rev. Lett.}\ }\textbf
  {\bibinfo {volume} {103}},\ \bibinfo {pages} {160502} (\bibinfo {year}
  {2009})}\BibitemShut {NoStop}%
\bibitem [{\citenamefont {Shanahan}\ \emph {et~al.}(2018)\citenamefont
  {Shanahan}, \citenamefont {Chenu}, \citenamefont {Margolus},\ and\
  \citenamefont {del Campo}}]{del_campo_qsl}%
  \BibitemOpen
  \bibfield  {author} {\bibinfo {author} {\bibfnamefont {B.}~\bibnamefont
  {Shanahan}}, \bibinfo {author} {\bibfnamefont {A.}~\bibnamefont {Chenu}},
  \bibinfo {author} {\bibfnamefont {N.}~\bibnamefont {Margolus}},\ and\
  \bibinfo {author} {\bibfnamefont {A.}~\bibnamefont {del Campo}},\ }\href
  {https://doi.org/10.1103/PhysRevLett.120.070401} {\bibfield  {journal}
  {\bibinfo  {journal} {Phys. Rev. Lett.}\ }\textbf {\bibinfo {volume} {120}},\
  \bibinfo {pages} {070401} (\bibinfo {year} {2018})}\BibitemShut {NoStop}%
\bibitem [{\citenamefont {Okuyama}\ and\ \citenamefont
  {Ohzeki}(2018)}]{PhysRevLett.120.070402}%
  \BibitemOpen
  \bibfield  {author} {\bibinfo {author} {\bibfnamefont {M.}~\bibnamefont
  {Okuyama}}\ and\ \bibinfo {author} {\bibfnamefont {M.}~\bibnamefont
  {Ohzeki}},\ }\href {https://doi.org/10.1103/PhysRevLett.120.070402}
  {\bibfield  {journal} {\bibinfo  {journal} {Phys. Rev. Lett.}\ }\textbf
  {\bibinfo {volume} {120}},\ \bibinfo {pages} {070402} (\bibinfo {year}
  {2018})}\BibitemShut {NoStop}%
\bibitem [{\citenamefont {Caneva}\ \emph {et~al.}(2009)\citenamefont {Caneva},
  \citenamefont {Murphy}, \citenamefont {Calarco}, \citenamefont {Fazio},
  \citenamefont {Montangero}, \citenamefont {Giovannetti},\ and\ \citenamefont
  {Santoro}}]{Caneva_PRL2009}%
  \BibitemOpen
  \bibfield  {author} {\bibinfo {author} {\bibfnamefont {T.}~\bibnamefont
  {Caneva}}, \bibinfo {author} {\bibfnamefont {M.}~\bibnamefont {Murphy}},
  \bibinfo {author} {\bibfnamefont {T.}~\bibnamefont {Calarco}}, \bibinfo
  {author} {\bibfnamefont {R.}~\bibnamefont {Fazio}}, \bibinfo {author}
  {\bibfnamefont {S.}~\bibnamefont {Montangero}}, \bibinfo {author}
  {\bibfnamefont {V.}~\bibnamefont {Giovannetti}},\ and\ \bibinfo {author}
  {\bibfnamefont {G.~E.}\ \bibnamefont {Santoro}},\ }\href
  {https://doi.org/10.1103/PhysRevLett.103.240501} {\bibfield  {journal}
  {\bibinfo  {journal} {Phys. Rev. Lett.}\ }\textbf {\bibinfo {volume} {103}},\
  \bibinfo {pages} {240501} (\bibinfo {year} {2009})}\BibitemShut {NoStop}%
\bibitem [{\citenamefont {Wauters}\ \emph {et~al.}(2020)\citenamefont
  {Wauters}, \citenamefont {Mbeng},\ and\ \citenamefont
  {Santoro}}]{Wauters_PRA2020}%
  \BibitemOpen
  \bibfield  {author} {\bibinfo {author} {\bibfnamefont {M.~M.}\ \bibnamefont
  {Wauters}}, \bibinfo {author} {\bibfnamefont {G.~B.}\ \bibnamefont {Mbeng}},\
  and\ \bibinfo {author} {\bibfnamefont {G.~E.}\ \bibnamefont {Santoro}},\
  }\href {https://doi.org/10.1103/PhysRevA.102.062404} {\bibfield  {journal}
  {\bibinfo  {journal} {Phys. Rev. A}\ }\textbf {\bibinfo {volume} {102}},\
  \bibinfo {pages} {062404} (\bibinfo {year} {2020})}\BibitemShut {NoStop}%
\bibitem [{\citenamefont {Christandl}\ \emph {et~al.}(2005)\citenamefont
  {Christandl}, \citenamefont {Datta}, \citenamefont {Dorlas}, \citenamefont
  {Ekert}, \citenamefont {Kay},\ and\ \citenamefont
  {Landahl}}]{Christandl_PRA2005}%
  \BibitemOpen
  \bibfield  {author} {\bibinfo {author} {\bibfnamefont {M.}~\bibnamefont
  {Christandl}}, \bibinfo {author} {\bibfnamefont {N.}~\bibnamefont {Datta}},
  \bibinfo {author} {\bibfnamefont {T.~C.}\ \bibnamefont {Dorlas}}, \bibinfo
  {author} {\bibfnamefont {A.}~\bibnamefont {Ekert}}, \bibinfo {author}
  {\bibfnamefont {A.}~\bibnamefont {Kay}},\ and\ \bibinfo {author}
  {\bibfnamefont {A.~J.}\ \bibnamefont {Landahl}},\ }\href
  {https://doi.org/10.1103/PhysRevA.71.032312} {\bibfield  {journal} {\bibinfo
  {journal} {Phys. Rev. A}\ }\textbf {\bibinfo {volume} {71}},\ \bibinfo
  {pages} {032312} (\bibinfo {year} {2005})}\BibitemShut {NoStop}%
\bibitem [{\citenamefont {Kolodrubetz}\ \emph {et~al.}(2017)\citenamefont
  {Kolodrubetz}, \citenamefont {Sels}, \citenamefont {Mehta},\ and\
  \citenamefont {Polkovnikov}}]{KOLODRUBETZ20171}%
  \BibitemOpen
  \bibfield  {author} {\bibinfo {author} {\bibfnamefont {M.}~\bibnamefont
  {Kolodrubetz}}, \bibinfo {author} {\bibfnamefont {D.}~\bibnamefont {Sels}},
  \bibinfo {author} {\bibfnamefont {P.}~\bibnamefont {Mehta}},\ and\ \bibinfo
  {author} {\bibfnamefont {A.}~\bibnamefont {Polkovnikov}},\ }\href
  {https://doi.org/https://doi.org/10.1016/j.physrep.2017.07.001} {\bibfield
  {journal} {\bibinfo  {journal} {Physics Reports}\ }\textbf {\bibinfo {volume}
  {697}},\ \bibinfo {pages} {1} (\bibinfo {year} {2017})}\BibitemShut {NoStop}%
\bibitem [{\citenamefont {Wurtz}\ and\ \citenamefont
  {Love}(2022)}]{Wurtz2022counterdiabaticity}%
  \BibitemOpen
  \bibfield  {author} {\bibinfo {author} {\bibfnamefont {J.}~\bibnamefont
  {Wurtz}}\ and\ \bibinfo {author} {\bibfnamefont {P.~J.}\ \bibnamefont
  {Love}},\ }\href {https://doi.org/10.22331/q-2022-01-27-635} {\bibfield
  {journal} {\bibinfo  {journal} {{Quantum}}\ }\textbf {\bibinfo {volume}
  {6}},\ \bibinfo {pages} {635} (\bibinfo {year} {2022})}\BibitemShut {NoStop}%
\end{thebibliography}%

\end{document}